%% file: ft_dls.tex
\documentclass[12pt]{article}
\usepackage[ruled,vlined]{algorithm2e}
\usepackage[noend]{algpseudocode}
\usepackage{multirow}
\usepackage{booktabs} 
\usepackage{subfig}
\usepackage{adjustbox}
\usepackage{dblfloatfix}

\usepackage{authblk}
\usepackage{lscape}

\usepackage{cite}
\usepackage{threeparttable}
\usepackage{amsmath,amssymb,amsfonts}
\usepackage{algorithmicx}
\usepackage{graphicx}
\usepackage{textcomp}
\usepackage[dvipsnames]{xcolor}
\def\BibTeX{{\rm B\kern-.05em{\sc i\kern-.025em b}\kern-.08em
		T\kern-.1667em\lower.7ex\hbox{E}\kern-.125emX}}

\definecolor{ForestGreen}{RGB}{63,142,38}
\definecolor{UnibasMint}{RGB}{30,165,165}

\definecolor{findNeighbors}{HTML}{0000FF}
\definecolor{Density}{HTML}{000075}
\definecolor{IAD}{HTML}{3791FF}
\definecolor{Divv}{HTML}{0B60FF}
\definecolor{Momentum}{HTML}{69CBFF}
\definecolor{Gravity}{HTML}{00AA00}
\definecolor{Communication}{HTML}{FF0000}
\definecolor{openMP}{HTML}{FE00FF}

\usepackage[normalem]{ulem}

\newcommand{\discuss}[1]{{\color{black}#1}}
\newcommand{\ali}[1]{{\color{black}#1}}

\newcommand{\dlbTool}{\mbox{\textit{DLS4LB}}}
\newcommand{\rdlb}{\mbox{\textit{rDLB}}}

\newcommand{\tl}{\mbox{thread-level}}
\newcommand{\pl}{\mbox{process-level}}

\newcommand{\cut}[1]{}

\begin{document}
 
\title{\rdlb{}: A Novel Approach for Robust Dynamic Load Balancing of Scientific Applications with Parallel Independent Tasks}

\author{Ali Mohammed}
\author{Aur{\'e}lien Cavelan}
\author{Florina M. Ciorba}	
\affil{Department of Mathematics and Computer Science\\
	University of Basel, Switzerland\\}
 
\renewcommand\Authands{ and }

\maketitle
\clearpage

\tableofcontents
\clearpage

\input{0}
\input{1}
\input{2}
\input{3}

\input{4}
\input{6}

\bibliographystyle{ieeetr}
\bibliography{citedatabase}
\end{document}

%% file: 0.tex
\begin{abstract}
\label{sec:abs}
Scientific applications often contain large and computationally-intensive parallel loops.
Dynamic loop self-scheduling~(DLS) is used to achieve a balanced load execution of such applications on high performance computing~(HPC) systems.
Large HPC systems are vulnerable to processors or node failures and perturbations in the availability of resources. 
Most self-scheduling approaches do not consider fault-tolerant scheduling or depended on failure or perturbation detection and react by rescheduling failed tasks.
In this work, a robust dynamic load balancing~(\rdlb{}) approach is proposed for the robust self-scheduling of independent tasks.
The proposed approach is proactive and does not depend on failure or perturbation detection.
The theoretical analysis of the proposed approach \discuss{shows that it is linearly scalable and its cost decrease quadratically by increasing the system size}. 
\rdlb{} is integrated into an MPI DLS library to evaluate its performance experimentally with two \mbox{computationally-intensive} scientific applications.
Results show that \rdlb{} enables the tolerance of up to ($P-1$)~processor failures, where $P$ is the number of processors executing an application.
In the presence of perturbations, \rdlb{} boosted the robustness of DLS techniques up to $30$ times and decreased application execution time up to $7$ times compared to their counterparts without \rdlb{}. 

\end{abstract}


\textbf{keywords.}
Dynamic loop self-scheduling, independent task scheduling, Robustness, Fault tolerance, High performance computing, Fail-stop errors, Perturbations

%% file: 1.tex
\section{Introduction}
\label{sec:intro}

Scientific applications are often large and computationally-intensive.
Parallel loop iterations are a vast resource of parallelism in these applications, in the form of independent tasks, where a loop iteration is a task. 
Large scientific applications are executed on high performance computing~(HPC) systems to fulfill their computational need.
Due to the irregularity of parallel tasks or HPC system characteristics, the performance of scientific applications may degrade as a result of load imbalance.
Dynamic loop self-scheduling (DLS) is used to balance the execution of parallel loops within \mbox{computationally-intensive} applications.
As HPC systems become larger to accommodate the computational needs of scientific applications, they become more vulnerable to faults and perturbations.
In fact, observed failures grow proportional to the number of processor sockets in a system~\cite{schroeder2007understanding}.
In addition, as smaller transistors are used for high integration in modern and future systems, they more prone to faults and failures due to their low operating voltage and wear-out. 
Besides failures, perturbations in the availability of the processing elements (PEs), network bandwidth, or network latency could also degrade applications performance.
These perturbations may occur due to other applications that share the same (network) resources, operating system interference, or transient malfunction~\cite{Mohammed:2018c}.

Current DLS approaches, such as \textit{DLB\_tool}~\cite{carino2007tool} and its extension \dlbTool{}~\cite{DLS4LB} do not address failures.
An approach for fault tolerant DLS where failed tasks are rescheduled on other PEs was evaluated in simulation~\cite{dlsmsg}.
However, this approach depends on detecting PE failures and reacting afterward by rescheduling failed tasks. 
For scheduling under perturbations, reinforced learning or simulation was used to learn~\cite{Boulmier:2017a} or predict~\cite{Mohammed:2018c} the DLS technique that would improve the performance given certain measured or predicted perturbations during execution.   

In this work, robust dynamic load balancing~(\rdlb{}) is introduced for the robust self-scheduling of independent tasks.
The proposed \rdlb{} approach is proactive rather than reactive and does not depend or need any failure or perturbation detection or measurement mechanism.
\rdlb{} is used to extend the MPI \dlbTool{} DLS library to enable robust scheduling in the presence of failures and perturbations.
The FePIA procedure~\cite{ali2004measuring} is used to evaluate the robustness of the proposed approach during the execution of two computationally-intensive scientific applications in different failure and perturbation scenarios.

This work makes the following contributions:
(1)~The introduction of a robust dynamic load balancing~(\rdlb) approach that applies the self-scheduling concept; 
(2)~A lower bound of the expected performance with \rdlb{} and the cost in the case of a single PE failure through the theoretical analysis of the proposed approach;
(3)~The extension of the \dlbTool{} with the \rdlb{} approach to employ robust dynamic loop self-scheduling;
(4)~The evaluation of the proposed approach through native experiments and injecting failures and perturbation.
In addition, the FePIA procedure is applied to calculate the robustness of the proposed approach in the execution of two scientific applications under failures and perturbations.

The rest of this work is structured as follows. 
Section~\ref{sec:background} offers the required background for this work. 
In addition, a review of the most related works is presented.
The description, theoretical analysis and the extension of \dlbTool{} with \rdlb{} is described in Section~\ref{sec:ft-dls}.
The design of experiments, results, and the evaluation of the proposed approach is discussed in Section~\ref{sec:ex}.
Section~\ref{sec:conc} concludes this work and outlines future work. 


%% file: 2.tex
\section{Background and Related Work}
\label{sec:background}
An overview of the most successful dynamic loop scheduling (DLS) techniques that improve the performance of computationally-intensive scientific applications via load balancing is given. 
Also, relevant definitions of faults, errors, failures, fault tolerance, and robustness are provided. 
The most related work on fault-tolerant and robust scheduling is discussed thereafter.

\subsection{Background}
\label{subsec:background}
\textbf{Loop scheduling.} \ali{Loop iterations are assigned via loop scheduling} to PEs to achieve a balanced load execution with minimum overhead.
Loop scheduling techniques are divided into static and dynamic.
Static loop scheduling techniques divide and assign loop iterations \ali{to} PEs before the execution starts. 
The division of loop iterations \ali{nor} their assignment do not change \ali{during execution}.
Examples of static loop scheduling techniques are block, cyclic, and block-cyclic~\cite{li1993locality}. 
In this work, block scheduling is considered, \ali{denoted as STATIC,} where each PE is assigned a block (chunk) of loop iterations equal to the total number of loop iterations, $N$, divided by the number of PEs, $P$. 
DLS techniques assign a chunk of loop iterations to free and requesting PEs during execution via self-scheduling.
DLS techniques can be further divided into nonadaptive and adaptive techniques. 
Nonadaptive DLS techniques address the load imbalance \ali{caused by} problem or application characteristics, such as the variation of loop iterations execution times. 
Nonadaptive DLS techniques include self-scheduling~\cite{SS}~(SS), fixed-sized chunk~\cite{FSC}~(FSC), modified fixed-sized chunk~\cite{banicescu:2013:a}~(mFSC), guided self-scheduling~\cite{GSS}~(GSS), trapezoid self-scheduling~\cite{tzen1993trapezoid}~(TSS), factoring~\cite{FAC}~(FAC), weighted factoring~\cite{WF}~(WF), and random~\cite{Ciorba:2018}~(RAND).
SS assigns a single loop iteration at a time per PE request. 
Thus, it results in the maximum load balance and the maximum scheduling overhead.
SS represents one extreme, where the load balancing \ali{effect} and the scheduling overhead \ali{are at} maximum, \ali{whereas} STATIC represents the other extreme, \ali{where load balancing effect and the scheduling overhead are at minimum}.
FSC assigns loop iterations in chunks of fixed size, \ali{hence reducing} the scheduling overhead compared to SS. 
The chunk size depends on the scheduling overhead, $h$, and the standard deviation of the iterations execution time, $\sigma$.
mFSC alleviates the burden of determining $h$ and $\sigma$ and assigns a chunk size that results in a number of chunks that is similar to that of FAC (explained below).
GSS addresses the uneven starting times \ali{of PEs} and assigns chunks in decreasing sizes. 
The chunk \ali{sizes in GSS are calculated as} the number of the remaining loop iterations, $R$, divided by $P$.
TSS assigns chunks of decreasing sizes, similar to GSS. 
However, chunk sizes decrease linearly in TSS, \ali{which simplifies} the chunk calculation and reduces scheduling overhead.
FAC assigns chunks in batches to reduce the scheduling overhead.
FAC \ali{employs} probabilistic analysis of application characteristics to calculate batch sizes that maximize the probability of achieving a balanced load execution.
The batch size calculation depends on the mean of iterations execution times, $\mu$, and their standard deviation, $\sigma$. 
The chunk sizes are equal \ali{in} a batch, \ali{namely} the batch size divided by $P$.
When $\mu$ and $\sigma$ are not available, FAC is practically implemented by assigning half of the remaining loop iterations \ali{as} a batch, \ali{which is equally distributed to PEs on request}.
WF is similar to FAC, except that it addresses heterogeneous PEs.
In WF, each PE is assigned a relative weight that is fixed during execution. 
Each PE is assigned a chunk \ali{from the current} batch relative to its weight.
In this work, the practical implementations of FAC and WF are used.
RAND employs the uniform distribution to arrive at a randomly calculated chunk size between an upper and a lower bound. 
The randomly calculated chunk size is bounded by $ N/ (100 \times P) \le chunk\_size \le N/(2 \times P)$~\cite{Ciorba:2018}.

Adaptive DLS techniques measure the performance during execution and adapt \ali{their} chunk calculation accordingly to address the load imbalance due to systemic characteristics, such as non-uniform memory access (NUMA) delays and perturbations during execution.  
Adaptive DLS techniques include adaptive \ali{weighted} factoring~\cite{AWF}~(AWF), and its variants~\cite{AWFBC}~AWF-B,~AWF-C,~AWF-D, and~AWF-E and adaptive factoring~\cite{AF}~(AF), among others.
AWF adapts the relative PE weights during execution according to their performance.
It is designed for \mbox{time-stepping} applications. 
It measures the performance of PEs during previous \mbox{time-steps} and updates the PEs relative weights after each \mbox{time-step} to balance the load according to the computing system's \ali{present} state.
\mbox{AWF-B} relieves the \mbox{time-stepping} \ali{requirement} to learn the PE weights.
It learns the PE weights from their performance in previous batches instead of \mbox{time-steps}.
\mbox{AWF-C} is similar to \mbox{AWF-B}, however, the PE weights are updated after the execution of each chunk, instead of batch.
\mbox{AWF-D} is similar to \mbox{AWF-B}, where the scheduling overhead (time taken to assign a chunk of loop iterations) is taken into account in the weight calculation. 
\mbox{AWF-E} is similar to \mbox{AWF-C}, and takes into account also the scheduling overhead, similar to \mbox{AWF-D}.
AF is \ali{also} based on FAC.
\ali{However,} it measures the performance of PEs to learn \ali{the} $\mu$ and $\sigma$ per PE during execution.

\textbf{DLS in Scientific Applications.}
Several scientific applications use dynamic loop scheduling to improve their performance on HPC systems.
DLS can be used at the \pl{} or the \tl{}.
For example at the \pl{}, SS, FAC, AWF, and AF were used in heat diffusion application on an unstructured grid~\cite{banicescu2001load}.
Also, FSC, WF, and AF were used to balance the load of the shadow process in decision support systems for solar energy potential~\cite{7830688}.
DLS techniques have also been used to balance the load of automatic quadrature routines~\cite{AWFBC}, N-Body simulations~\cite{fractiling}, and parallel spin image algorithm (PSIA)~\cite{Eleliemy:2017b}.
At the \tl{}, SS, TSS, GSS, FAC, and WF were used to balance the load in kernels of several benchmarks such as RODINIA suite, NAS OpenMP suite, and SPEC OpenMP~\cite{Ciorba:2018}.

\textbf{Robustness.}
A \emph{fault} is a sudden malfunction that occurs in the computing system, such as a bit flip or incorrect control signal.
A fault can lead to an \emph{error} when a faulty unit is used in calculations. 
Errors could be fatal, which result in a \emph{fail-stop failure}, meaning that the system will cease its normal service, or they could be silent or dormant errors, such as \emph{silent data corruptions} (SDC)~\cite{Fiala:2012:DCS:2388996.2389102}. 
In this work, fail-stop failures of computing cores, nodes, or network links that render certain nodes are unreachable are considered and denoted as \emph{failures} in the rest of this text.
Besides failures, the available computing speed of a PE or the network latency or bandwidth could change during applications execution due to resource sharing or temporary (instantaneous) malfunctions.
These variations in the computing speed or network latency or bandwidth are referred to as \emph{perturbations} and can degrade the application performance running on large scale HPC system.
Maintaining ``correct'' operation in the presence of failures is denoted as \emph{fault tolerance}, whereas \emph{robustness} denotes the maintenance of certain desired system characteristics despite fluctuations in the behavior of its components or its environment~\cite{ali2004measuring}. 
Robustness of DLS techniques to variations in the computing unit availability is denoted as \emph{flexibility}~\cite{srivastava:2010} whereas the robustness against fail-stop PE failures is denoted as \emph{resilience}.  

\textbf{Failures in HPC systems.}
HPC systems are large and complex that are built with quality components.
However, due to the large number of components in these systems, failures are inevitable.
For example, ASC~Q system at Los Alamos National Laboratory had on average~$26.1$~processors failures per week~\cite{michalak2005predicting}.
Also, the mean time between failures (MTTF) for BlueGene and Titan were found to be $7.9$~days~\cite{bergman2008exascale} and $22.78$~hours~\cite{ni2016mitigation}, respectively.
In fact, failure rates of a system grow proportional to the number of processors sockets in a system~\cite{schroeder2007understanding}.
The study of failures in large scale systems revealed that there is a correlation between the failure rate of a system and the type and intensity of the workload execution on it~\cite{schroeder2010large}.
In addition, failures exhibit temporal correlation and are spatially correlated as well, especially for network root-caused failures.
Extrapolating the current failure rates to Exascale systems would result in MTTF of $24$~minutes, and if the resiliency of the components is assumed to be improved by a factor of $10$, this would result in a failure every $4$~hours~\cite{bergman2008exascale}.

\subsection{Related Work}
\label{subsec:related_work}
The flexibility of nonadaptive DLS techniques on distributed heterogeneous HPC systems was studied~\cite{DBLP:conf/parco/Garcia-Gonzalez17}.
Also, the flexibility of adaptive and nonadaptive DLS techniques has been studied~\cite{sukhija:2014:a} and machine learning was used to create a portfolio of DLS techniques flexibility. 
However, none of the work above considered sources of perturbations other than the variability of PEs availabilities, such as unpredictable network latency or bandwidth. 
RUMR~\cite{yang2003rumr} was introduced as a robust scheduling method to unpredictable task execution time and unpredictable communication time. 
Multi-objective evolutionary algorithm and a robust version of HEFT~\cite{topcuoglu2002performance} were introduced for the robust scheduling of tasks with uncertain computation and communication time~\cite{canon2010evaluation}.
Static scheduling methods such as robust HEFT or evolutionary algorithm optimization can not adapt to unpredictable perturbations and failures that might occur during execution.
Dynamic selection of the most robust DLS technique using reinforced learning was used introduced~\cite{Boulmier:2017a} and simulation was used to predict and select the best scheduling techniques under unpredictable perturbations during execution to achieve the best application performance~\cite{Mohammed:2018c}.
DLS selection methods above are reactive rather than proactive and depend on detecting perturbations during execution either by measuring or predicting. 
Also, non-preemptive scheduling could result in poor performance in certain cases due to frequent change of the selected DLS.

A fault-tolerant approach for DLS was introduced and studied using simulation~\cite{dlsmsg}, where failed tasks were rescheduled dynamically to working PEs.
Different numbers of failed PEs were simulated that represent $12.5\%$, $25\%$, and $50\%$ of the total PEs in the system and the resilience of different DLS techniques were measured based on the number of tasks that needed to be rescheduled in each case. 
The above work introduced fault tolerance to the DLS techniques. However, it is a reactive approach and depends on detecting a PE failure and then reacting by rescheduling the failed tasks on working PEs.
Moreover, the failure detection method was not explicitly clear in this work, as it was only studied in a simulation.

Fault tolerant \mbox{self-scheduling}~\cite{wang2012fault} (FTSS) was introduced for shared memory systems.
FTSS incorporate \mbox{work-stealing} with \mbox{self-scheduling} for fault tolerance.
An idle thread will steal work from other loaded/failed thread after all loop iterations are already scheduled. 
This is different from the proposed \rdlb{}, where PEs are assigned work from a central work queue that contain all unfinished tasks and not steal from a certain victim PE.
Also, work stealing and victim selection depend on failure detection in FTSS, which was not discussed in this related work.

Fault tolerant \mbox{work-stealing}~\cite{wang2013work}~(FTWS) was presented for distributed memory systems. 
Tasks are duplicated and executed on two different PEs to detect transient and permanent PE failures.
Failed tasks are pushed to the faulty task queue, where PEs executes its tasks as soon as they finish their original work queue.
An approach was proposed for enhancing the flexibility of scheduling of a bag of tasks on compute grids~\cite{da2003trading}.
Only perturbations in the computing resources and variations in task sizes were considered and the proposed method was evaluated in simulation.
In contrast, the present work considers both failures and perturbations in the computing resources and the network and evaluates the \rdlb{} via native experiments using \mbox{MPI-based} library.
Similar robust scheduling approach to restart failed tasks can be configured for Apache Spark and Hadoop YARN~\cite{Yarn}.

This work proposes a robust approach to DLS, namely: \rdlb{}, that tolerates up to $P-1$ failures, where $P$ is the number of PEs used by the application. 
In addition, the proposed approach achieves improved performance in cases of severe PE availability or network perturbations.
The proposed approach is proactive and does not depend or need any failures or perturbations detection mechanism.
The proposed approach is integrated into a state-of-the-art DLS MPI library, namely: \dlbTool{}~\cite{DLS4LB}, and its performance is evaluated by theoretical modeling and in native experiments.
%

%% file: 3.tex
\section{Proposed Approach: \rdlb{}} 
\label{sec:ft-dls}

A Robust DLS approach is proposed herein, where processes (PEs and nodes) failures are tolerated. 
In the proposed approach, each loop iteration is flagged as \texttt{Unscheduled}, or \texttt{Scheduled}, or \texttt{Finished}. 
At the beginning of the execution, all loop iterations are \texttt{Unscheduled}. 
Loop iterations are scheduled via DLS techniques to free and requesting PEs, and changes their flag from \texttt{Unscheduled} to \texttt{Scheduled}.
In normal non-robust DLS execution, the scheduling operations end when all loop iterations are scheduled to PEs. 
In the proposed approach, scheduling continues after all loop iterations are scheduled, and \emph{reschedule scheduled and unfinished} loop iterations, that has their flag set to \texttt{Scheduled}. 
The \emph{key} idea of the proposed approach is to \emph{leverage the idle time processes wait at the end of the execution for robustness}.
The execution completes when all loop iterations are executed and have their flag set to \texttt{Finished}.
\begin{figure}[htbp]
	\centering
	\includegraphics[clip, trim=0cm 0cm 0cm 0cm, scale=0.95]{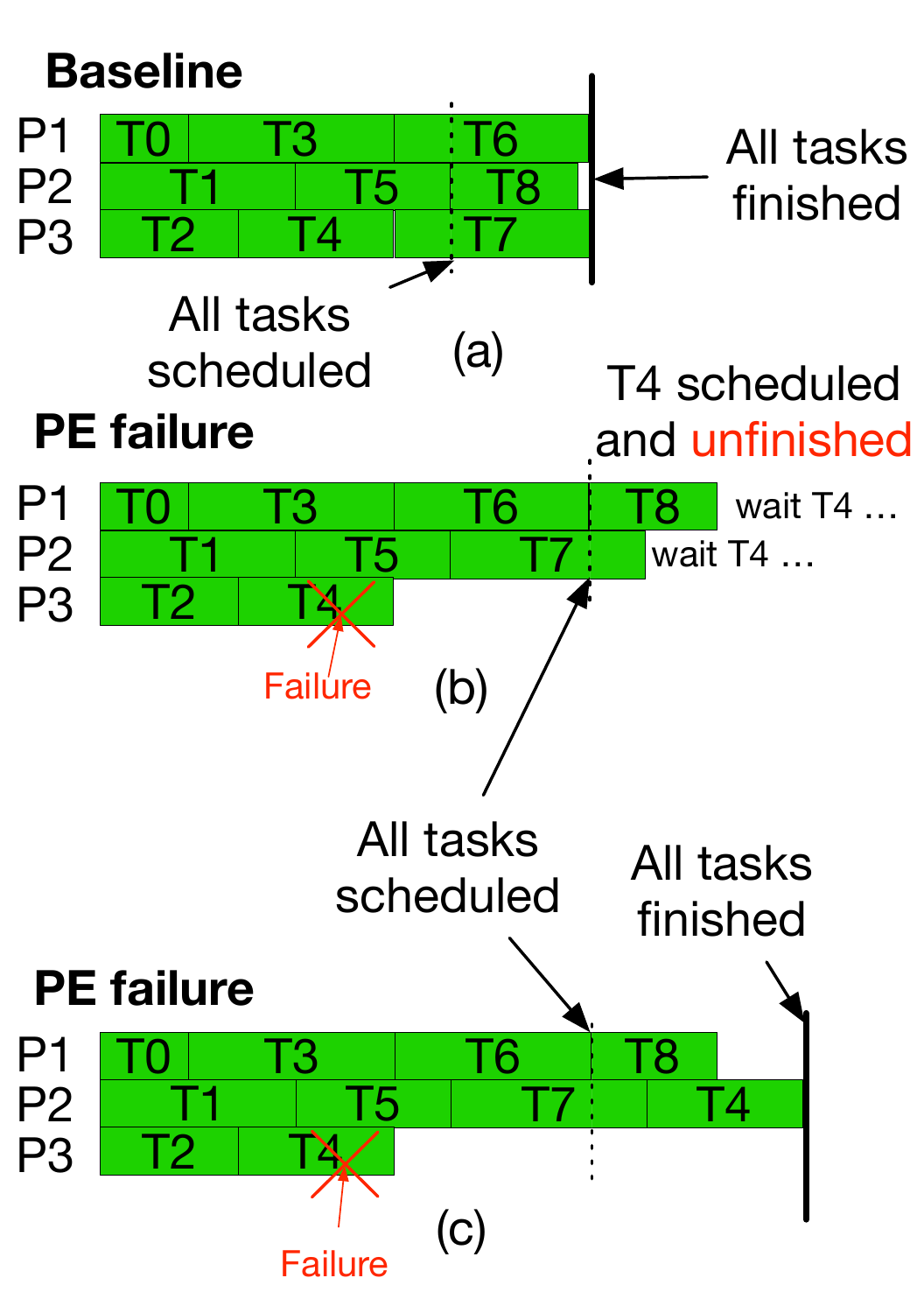}%
	\caption{A conceptual illustration of the execution of $9$ tasks on $3$ PEs with and without the proposed robust \rdlb{} approach in the presence of failures. Sub-figure~(a) shows the execution with SS in the case of no failures, (b)~the execution with SS in the case of one failure, and (c)~the execution with SS and \rdlb{} in the case of one failure.}
	\label{fig:approach_ft}
\end{figure}

\figurename{~\ref{fig:approach_ft}} illustrates the execution with and without the proposed robust DLS approach in the presence of fail-stop failure or perturbations.
In case of PE failure, \texttt{T4} will never be executed because from scheduling it was already scheduled on $P3$ which failed later, and the execution would wait indefinitely for $P3$ to finish \texttt{T4} and complete the execution. 
With \rdlb{}, after all tasks are scheduled, and $P2$ becomes available, and requests work, the first \emph{scheduled and unfinished} task, \texttt{T4} is assigned to it, and the execution completes as soon as all tasks are \texttt{Finished}.
\begin{figure}[]
	\centering
	\includegraphics[clip, trim=0cm 0cm 0cm 0cm, scale=0.95]{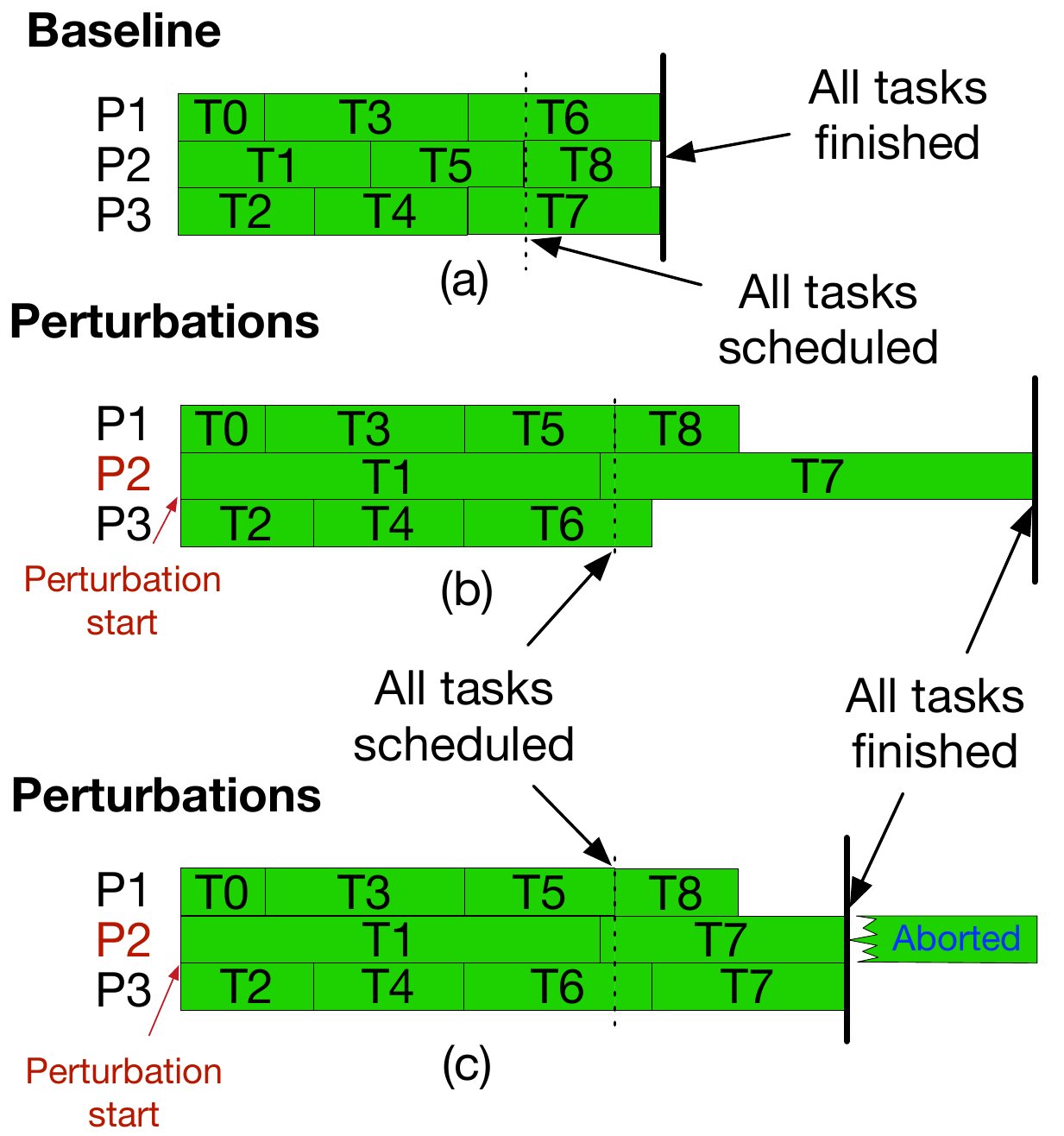}%
	\caption{A conceptual illustration of the execution of $9$ tasks on $3$ PEs with and without the proposed robust \rdlb{} approach in the presence of perturbations. Sub-figure~(a) shows the execution with SS in the case of no perturbations, (b)~the execution with SS in the case of severe perturbations on P2, and (c)~the execution with SS and \rdlb{} in the case of severe perturbations on P2.}
	\label{fig:approach_pert}
\end{figure}

Similarly in the case of a severe perturbation, represented by \figurename{~\ref{fig:approach_pert}}, tasks assigned to $P2$, the perturbed PE, takes much longer to complete execution. 
With the proposed robust approach, \texttt{T7} is rescheduled on $P3$ and execution completes earlier than that without \rdlb{}. 
The advantage of the proposed approach is that it acts \emph{proactively} and does not depend on any failure or perturbation mechanism.
Rescheduling of unfinished loop iterations or tasks does not entirely add to the execution time, as it overlaps with the idle time when all loop iterations are scheduled and PEs are waiting for the completion of all loop iterations.
The execution terminates as soon as either the original or the rescheduled tasks complete.

\subsection{Theoretical analysis }

Without taking into account the overhead of the scheduling algorithm and the communication overhead between PEs:
\begin{itemize}
\item Without failure, general case:
\begin{align*}
T &= \max\left(T_{P_1}, T_{P_2}, \ldots, T_{P_q}\right)
\end{align*}
with $T_{P_i} = \sum_{i=0}^{n_i} t_i$.



\item Without failure, all tasks equal and equally distributed:
	\begin{align*}
		T &= n \cdot t
	\end{align*}

\item With one failure only, all tasks equal and equally distributed:
\begin{align*}
\mathbb{E}_T &= \left(1-p_F^T\right) \cdot T + p_F^T \left(T + \sum_{i=0}^{n-1} \frac{1}{n} \cdot \frac{n-i}{q-1} \cdot t\right)  \\
&= T + p_F^T \left(\frac{t}{n}\sum_{i=0}^n \frac{n-i}{q-1}\right) \\
&= T + p_F^T \frac{t}{2}\frac{n+1}{q-1}
\end{align*}
With exponential failures:
\begin{align*}
\mathbb{E}_T &= T + (1-e^{-\lambda T}) \frac{t}{2}\frac{n+1}{q-1} \ .
\end{align*}
First-order approx:
\begin{align*}
\mathbb{E}_T &= T + \lambda T \frac{t}{2}\frac{n+1}{q-1} \ .
\end{align*}
Overhead:
\begin{align*}
\mathbb{H}_T &= 1-\frac{\mathbb{E}_T}{T}=\frac{\lambda t}{2}\frac{n+1}{q-1} \ .
\end{align*}
Overhead checkpointing:
\begin{align*}
\mathbb{H}_T^{C} &= \sqrt{2\lambda C} \ .
\end{align*}
FT-DLS is better than checkpointing when (first-order approx, i.e. $C << \frac{1}{\lambda}$):
\begin{align*}
C \geq = \frac{\lambda t^2}{8} \frac{(n+1)^2}{(q-1)^2} \ .
\end{align*}
\end{itemize}

\subsection{Implementation}
\begin{algorithm}[htbp]
	\caption{Integration of \rdlb{} with the \dlbTool{}}
	\label{algo:robust_DLS}
	\#include $<$mpi.h$>$\\
	{\color{blue} \#include ``DLS4LB.h''}\\
	
	MPI\_Init(\&argc, \&argv);\\
	MPI\_Comm\_size(MPI\_COMM\_WORLD, \&$P$);\\ 
	MPI\_Comm\_rank(MPI\_COMM\_WORLD, \&myid);\\
	
	/*Initialization*/ \\
    {\color{blue} rDLB\_enable = 1;} \\
	{\color{blue} MPI\_Comm\_set\_errhandler(MPI\_COMM\_WORLD, MPI\_ERRORS\_RETURN);}\\
	{\color{blue} DLS\_setup(MPI\_COMM\_WORLD, DLS\_info, rDLB\_enable);} \\
	{\color{blue} DLS\_startLoop (DLS\_info, $N$, DLS\_method, results\_data);}\\

	{\color{blue} \While{ {\color{blue} Not DLS\_terminated}}
		{
			{\color{blue} DLS\_startChunk(DLS\_info, assigned\_iters, size);}\\
			{\color{black} 
				{\color{blue} data = malloc(size);}\\
				Compute\_iterations(assigned\_iters, size, data);\\
			} 
			{\color{blue} DLS\_endChunk(DLS\_info, data);}\\
			free(data);\\
		} 
	} 
	DLS\_endLoop(DLS\_info);\\
	/*save results*/\\
	\dots\\
	/*End program immediately, kill all processes*/
	{\color{blue} MPI\_Abort(MPI\_COMM\_WORLD, -1);}  
\end{algorithm}	

The proposed \rdlb{} approach is used to extend the \dlbTool{} MPI scheduling library with robustness. 
The \dlbTool{} implements 13 DLS techniques (nonadaptive and adaptive) and uses a master-worker execution model.
Algorithm~\ref{algo:robust_DLS} shows how \dlbTool{} is used for robust scheduling.
The MPI error handler has to be changed to \texttt{MPI\_ERRORS\_RETURN} which reports MPI errors, rather than considering them fatal and ends the program immediately on MPI errors (default behavior).
Workers ask the master for work whenever they become free.
Upon completing a chunk of loop iterations, the worker sends the results to the master and asks for more work. 
Upon receiving the results of a chunk from a worker, the master marks the loop iterations previously assigned to the requesting worker as \texttt{Finished} and assigns the worker new chunk.
The \dlbTool{} is adjusted such that scheduling operations are continued until all loop iterations are completed.
When the master marks all iterations as \texttt{Finished}, it exits the \texttt{while} loop immediately, save all the results, and calls \texttt{MPI\_Abort} to kill all other process and end the execution. 
\texttt{MPI\_Abort} is called to ensure that the execution will terminate immediately as soon as all iterations are finished, and not hang indefinitely in a collective operation such as \texttt{MPI\_Finalize} due to failed processes.
Alternatively, implementations of fault-tolerant MPI such as User-Level Failure Mitigation~\cite{bland2012evaluation} (ULFM), could be used to detect and exclude failed processes from the communicator. 
However, as such fault-tolerant implementations are not yet part of the MPI standard, the \texttt{MPI\_Abort} approach is chosen instead, as it complies with the MPI standard and suffices the requirement of terminating the execution, even in case of failed processes.  
The current implementation of the proposed robust DLS works proactively and does not depend on or require fault detection. 
It does not add overhead to the application execution time, as the loop iterations are only rescheduled on free PEs when they are waiting for the completion of the execution after all loop iterations are scheduled. 
The execution is terminated as soon as all loop iterations are completed and reported to the master successfully.
A limitation of the current implementation is the master, being a single point of failure. However, work is planned to transform the \dlbTool{} to a decentralized execution model using decentralized DLS implementation approaches previously explored at the \tl{}~\cite{HPCS} and the \pl{}~\cite{Eleliemy:2019}.


%% file: 4.tex
\section{Experimental Evaluation}  
\label{sec:ex}

\subsection{Design of Experiments}
A summary of the experiments performed to evaluate \rdlb{} is presented in Table~\ref{tbl:exp}.
In practice, PE or node failures are often noticed in long executions over a large number of PEs. 
For practical reasons, in this work, short applications execution times and small system size are used to show the benefit of the proposed approach. \\
\begin{table}[htbp]
	\caption{Details used in the design of factorial experiments}
	\label{tbl:exp}
	\centering
	\begin{adjustbox}{max width=\textwidth}
		\begin{threeparttable}
			\begin{tabular}{@{}lll@{}}
				\toprule
				\textbf{Factors}             & \textbf{Values}                                                                                                                  & \textbf{Properties}                                                                                                                                                               \\ \midrule
		\multirow{2}{*}{\textbf{Applications}}                                                                                                                                      & PSIA                                                                                                                                                                                                                    & \begin{tabular}[c]{@{}l@{}}Low variability among iterations\\ $N = 20,000$\tnote{a}\end{tabular}                                                                                                                                                                                                                    \\ \cmidrule(l){2-3} 
		& Mandelbrot                                                                                                                                                                                                              & \begin{tabular}[c]{@{}l@{}}High variability among iterations\\ $N = 262,144$\end{tabular}                                                                                                                                                                                                                  \\ \midrule
		\multirow{3}{*}{\begin{tabular}[c]{@{}l@{}}\textbf{Loop} \\ \textbf{scheduling}\end{tabular}}                                                                                        & STATIC                                                                                                                                                                                                                  & Static                                                                                                                                                                                                                                                                                                   \\ \cmidrule(l){2-3} 
		& \begin{tabular}[c]{@{}l@{}}SS, FSC, mFSC, GSS, \\ TSS, FAC, WF\end{tabular}                                                                                                                                             & \begin{tabular}[c]{@{}l@{}}Dynamic nonadaptive\\ (with and without \rdlb{})\end{tabular}                                                                                                                                                                                                \\ \cmidrule(l){2-3} 
		& AWF-B, -C, -D, -E, AF                                                                                                                                                                                                   & \begin{tabular}[c]{@{}l@{}}Dynamic adaptive\\ (with and without \rdlb{})\end{tabular}                                                                                                                                                                                                   \\ \midrule
				{\textbf{\begin{tabular}[c]{@{}l@{}}Execution \\scenarios \end{tabular}}}	& Baseline   & No failures or perturbations  \\ \cmidrule(l){2-3} 
				\textbf{- Failures}          & \begin{tabular}[c]{@{}l@{}}one failure\\ P/2 failures\tnote{b}\\ P-1 failures\end{tabular}                                                & \begin{tabular}[c]{@{}l@{}} Assumptions: \\(1)~Fail-stop failures\\ (2)~Failed cores do not recover\\ (3)~Occur arbitrary during execution\end{tabular}                                        \\ \cmidrule(l){2-3} 
				\textbf{- Perturbations}     & \begin{tabular}[c]{@{}l@{}}PE perturbations\\ Network latency perturbations\\ Combined PE and latency \end{tabular} & \begin{tabular}[c]{@{}l@{}}All PEs on a single node slow down\\Delay all communications of a single node\\ Combined\end{tabular} \\ \midrule
				\textbf{\begin{tabular}[c]{@{}l@{}}Computing\\ system\end{tabular}}    & miniHPC                                                                                                                          & \begin{tabular}[c]{@{}l@{}}16 Dual socket Intel Broadwell nodes\\ 10 cores per socket        \end{tabular}    \\ \bottomrule
			\end{tabular}
			\begin{tablenotes}
				\item[a] $N$ is the total number of loop iterations.
				\item[b] $P$ is the total number of PEs.
			\end{tablenotes}
		\end{threeparttable}
	\end{adjustbox}
\end{table}

\noindent\textbf{Applications.} Two \mbox{computationally-intensive} applications are considered in this work, the parallel spin image algorithm~(PSIA)~\cite{psia} and the calculation of the Mandelbrot set~\cite{mandelbrot1980fractal}.
PSIA is a computer vision application that converts 3D objects into sets of 2D descriptors. 
Mandelbrot is a well-known mathematical kernel, and it is often used to evaluate the performance of scheduling techniques due to the high variation between its loop iterations execution times.\\
\noindent\textbf{Loop Scheduling.} The \dlbTool{} is employed for scheduling the loop iterations in both applications.
The \dlbTool{} is extended in this work with the \rdlb{} approach to enable robust scheduling as shown in Section~\ref{sec:ft-dls}.\\
\noindent\textbf{Injecting failures and perturbations.} To simulate failures, ranks make exit calls during the computation of the loop at arbitrary times during execution.
To show the benefit of \rdlb{} in tolerating a high number of failures,  experiments are performed with $1$, $P/2$, and $P-1$ failures, where $P$ is the number of PEs.
Perturbations in PE availability is performed by running a CPU burner simultaneously on the same PEs as the running applications.
For perturbations in network latency, MPI communication calls are intercepted through MPI profiling interface, and $10$~seconds delays are added for any communication to or from a specified node. 
As computationally-intensive applications are the focus of this work, perturbations in network bandwidth have almost no effect on application performance~\cite{Mohammed:2018c}. 
Therefore, perturbations in network bandwidth are not considered in this work.\\
\noindent\textbf{Computing system.} The evaluation experiments are conducted on miniHPC, a research and teaching cluster at the Department of Mathematics and Computer Science at the University of Basel, Switzerland. 
Sixteen nodes are used in the experiments, with 16 ranks per node, 8 ranks per socket, with a total of 256 ranks on 256 PEs. \\
\noindent\textbf{Evaluation of robustness.} The FePIA~\cite{ali2004measuring} procedure in applied to the performance results to evaluate the robustness of \rdlb{}.
The resilience, understood as robustness against failures is calculated as $\rho_{res}(\phi,\pi)$ where $\phi$ is the performance feature, i.e., parallel loop execution time $T_{par}$ and $\pi$ is the perturbation parameter, i.e., PE failures~\cite{srivastava:2010}. 
$\rho_{res}(\phi,\pi) = r_{DLS}/ r_{minDLS}$, where the robustness radius $r_{DLS} = T_{par}^{\pi} - T_{par}^{orig}$ and $r_{minDLS}$ is the minimum robustness radius for certain perturbation parameter $\pi$.
$T_{par}^{\pi}$ is the parallel execution time under perturbation $\pi$ and $T_{par}^{orig}$ is the parallel execution time in the baseline.
$\rho_{res}(\phi,\pi)$ is calculated for $\pi_1$, $\pi_2$, and $\pi_3$, which corresponds to one PE failure, $P/2$ PE failures, and $P-1$ PE failures, respectively.
Similarly, flexibility defined as robustness against perturbations, is calculated as $\rho_{flex}(\phi,\pi)$, where $\phi$ is the parallel loop execution time $T_{par}$ and $\pi$ is the perturbation parameter, i.e., PE perturbations, latency perturbations, and combined perturbations for the three perturbations cases considered in this work. 
A $\rho_{res}$ or $\rho_{flex}$ of $1$ denotes the most robust technique (the metric), and the robustness of other techniques are calculated with respect to the most robust one (how many times they are less robust than the most robust technique).  
%
\begin{figure*}[htbp]
	\begin{minipage}{\textwidth}
		\centering
		\subfloat[PSIA - failures]{%
			\includegraphics[clip, trim=0cm 0cm 0cm 0cm, scale=0.36]{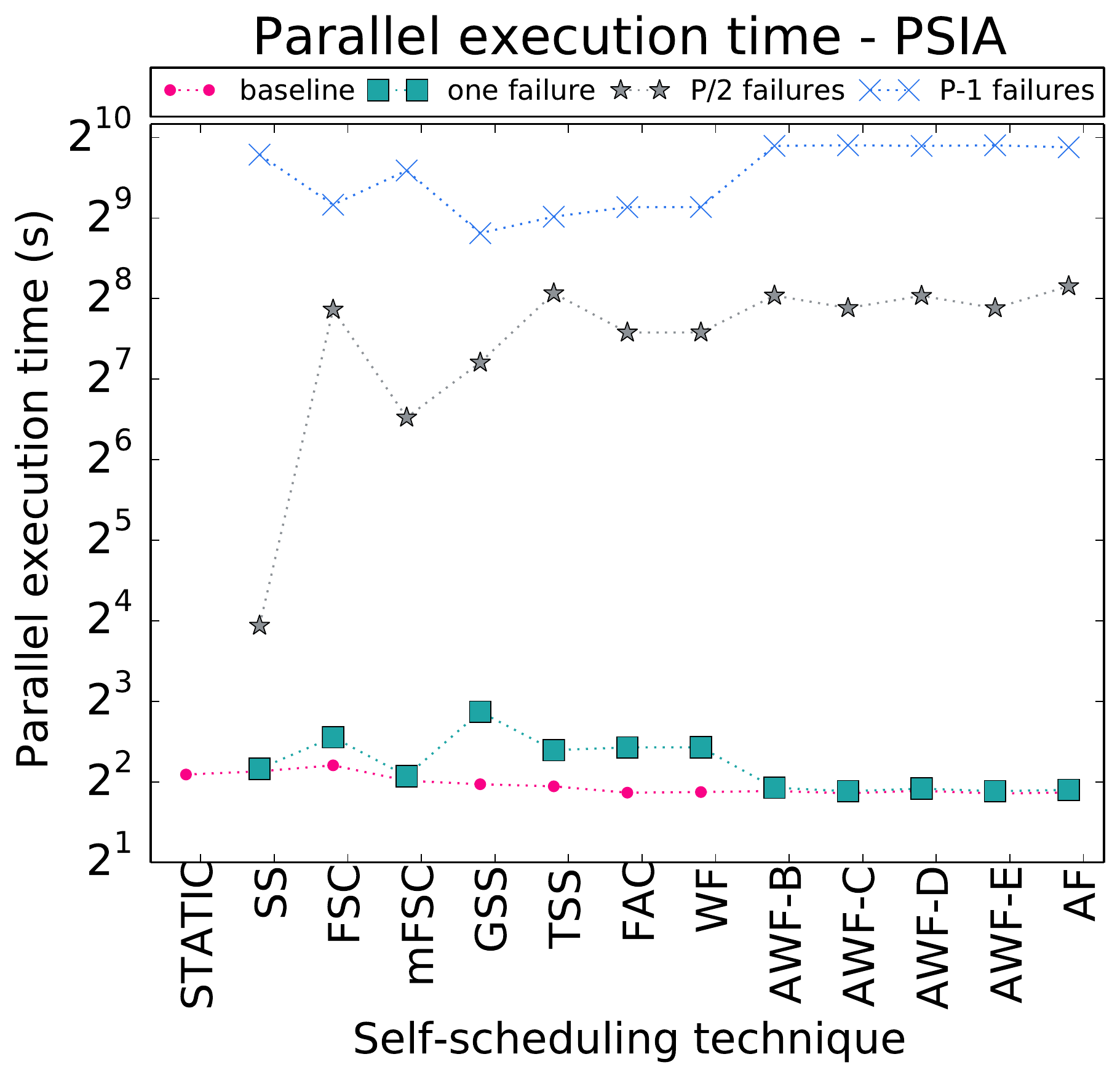}%
			\label{subfig:PSIA_ft}%
		} \hspace{0cm} 
		\subfloat[Mandelbrot - failures]{%
			\includegraphics[clip, trim=0cm 0cm 0cm 0cm, scale=0.36]{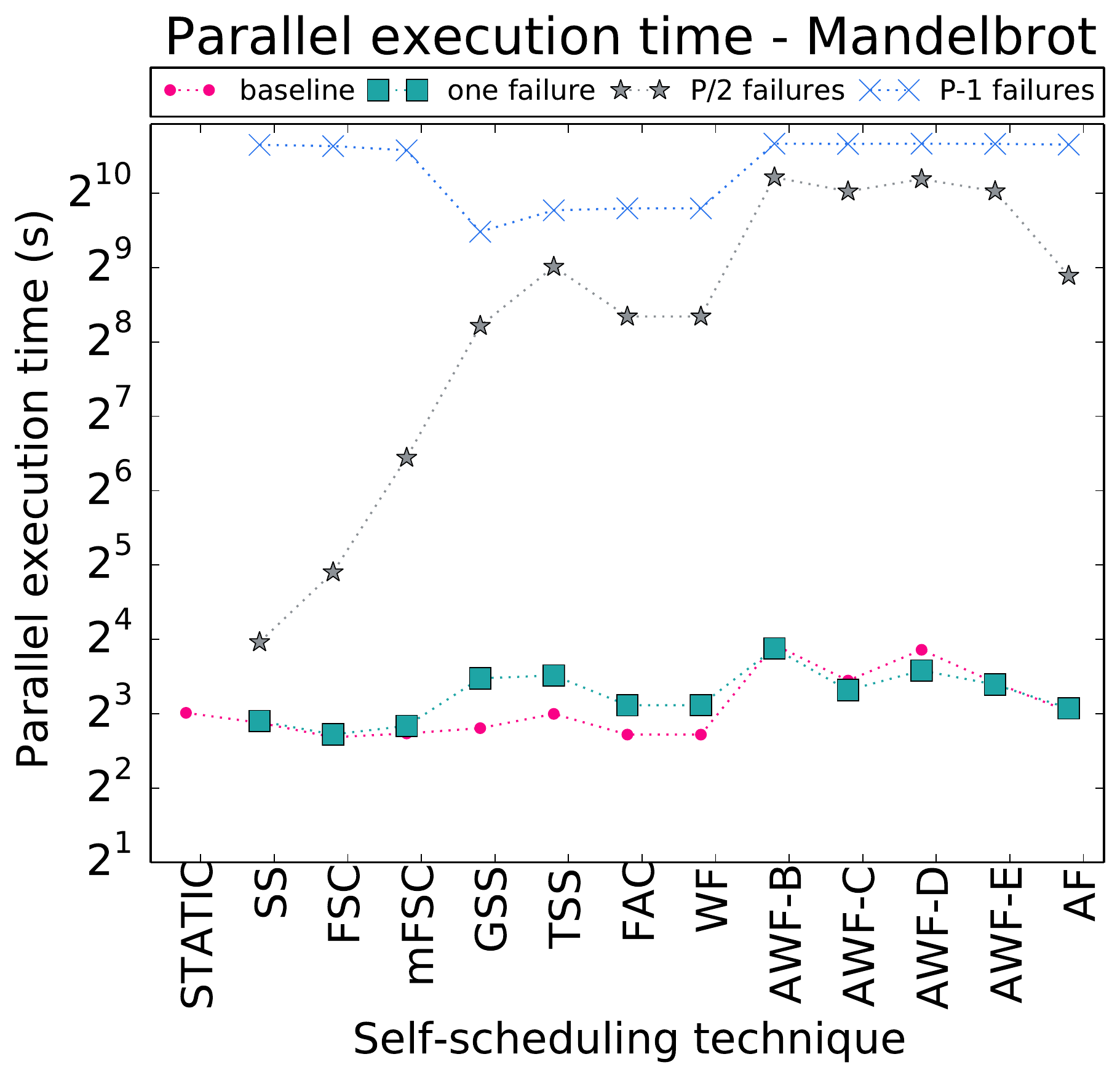}%
			\label{subfig:Mandel_ft}%
		} \\
		\subfloat[PSIA - perturbations ]{%
			\includegraphics[clip, trim=0cm 0cm 0cm 0cm, scale=0.36]{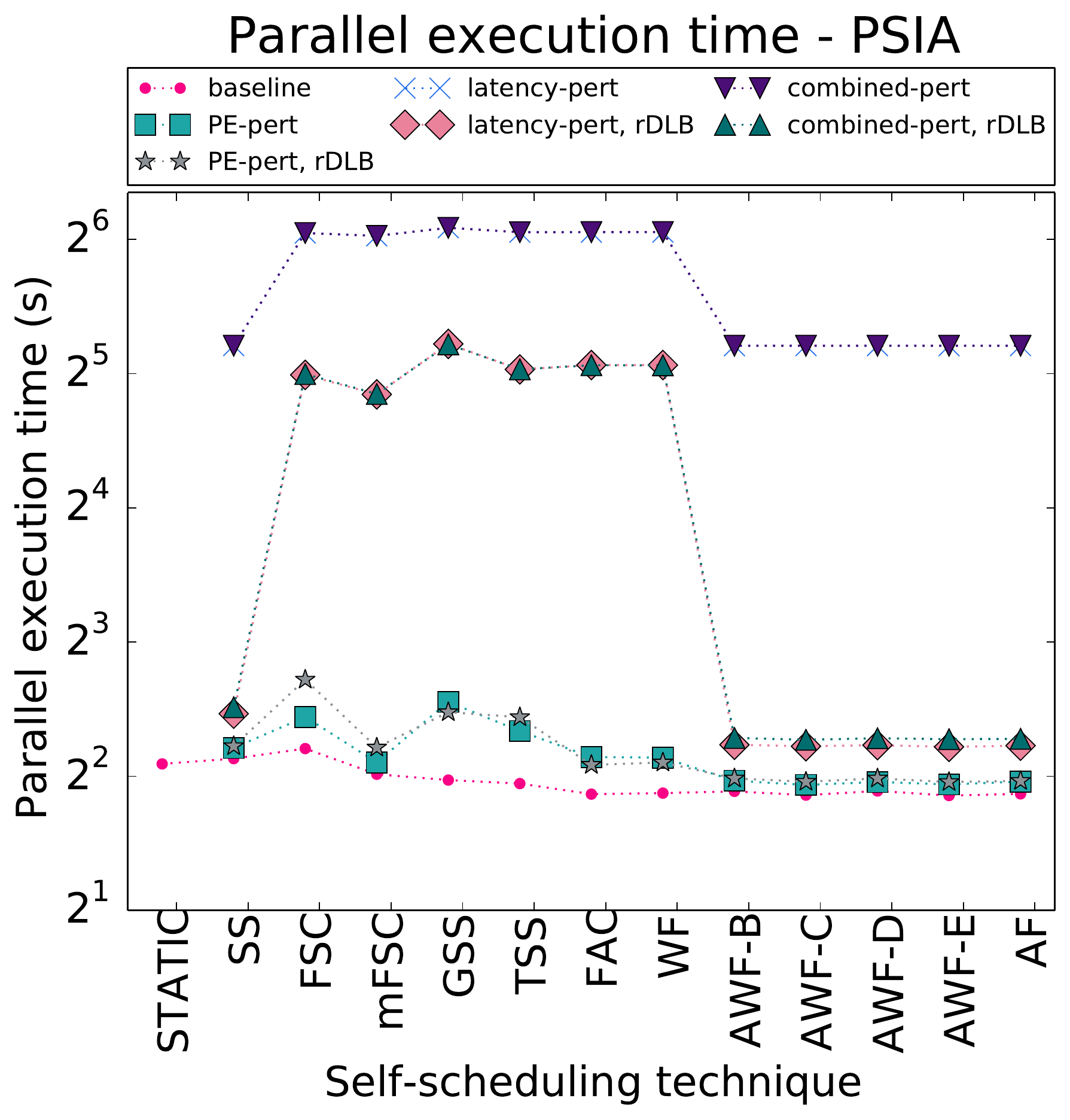}%
			\label{subfig:PSIA_pert}%
		} \hspace{0cm} 
		\subfloat[Mandelbrot - perturbations]{%
			\includegraphics[clip, trim=0cm 0cm 0cm 0cm, scale=0.36]{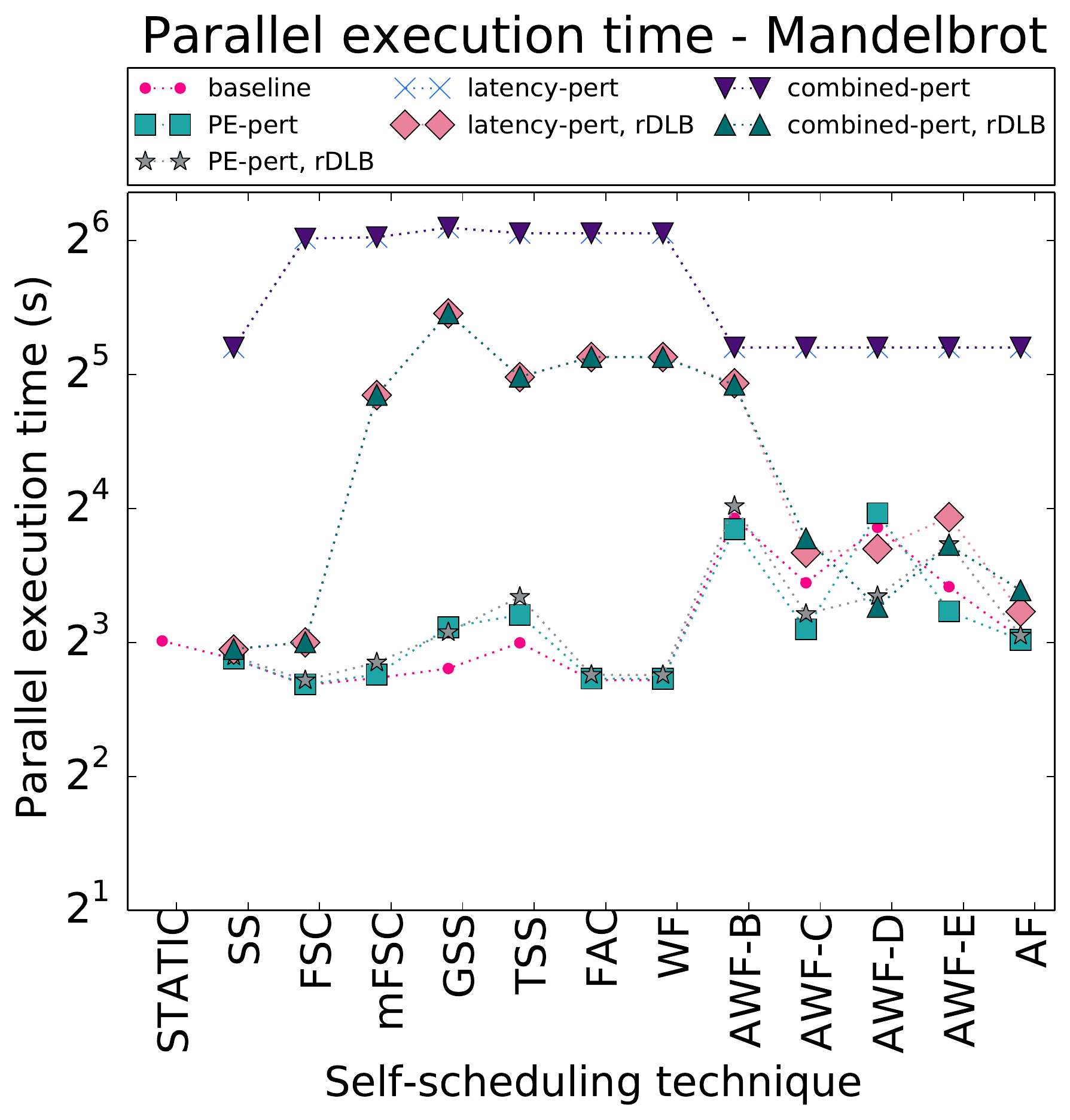}%
			\label{subfig:Mandel_pert}%
		} 	
		\caption{Performance of PSIA and Mandelbrot without and with the \rdlb{} under failures, PE perturbations, and latency perturbation scenarios on miniHPC with 256 cores. Execution with \rdlb{} tolerates up to $P-1$ failures. Also, \rdlb{} enhanced the performance of applications in the presence of perturbations with various DLS techniques by 7x with the adaptive DLS techniques and latency perturbations.}
		\label{fig:Tpar}
	\end{minipage}
\end{figure*}

\begin{figure*}[htbp]
	\begin{minipage}{\textwidth}
		\centering
		\subfloat{%
			\includegraphics[clip, trim=0cm 0cm 0cm 0cm, scale=0.24]{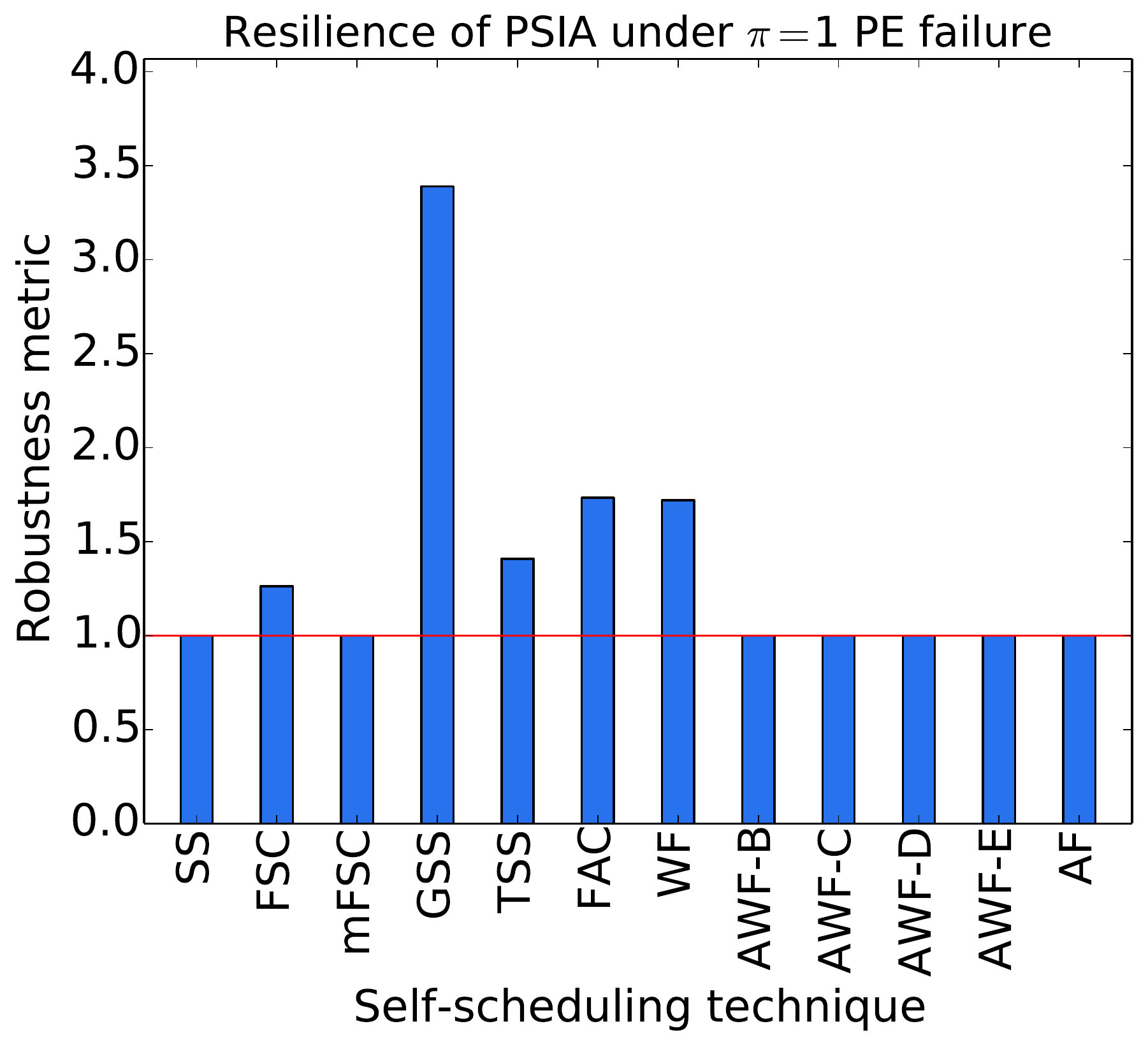}%
			\label{subfig:PSIA_ft_metric_1}%
		} \hspace{0cm} 
		\subfloat{%
		\includegraphics[clip, trim=0cm 0cm 0cm 0cm, scale=0.24]{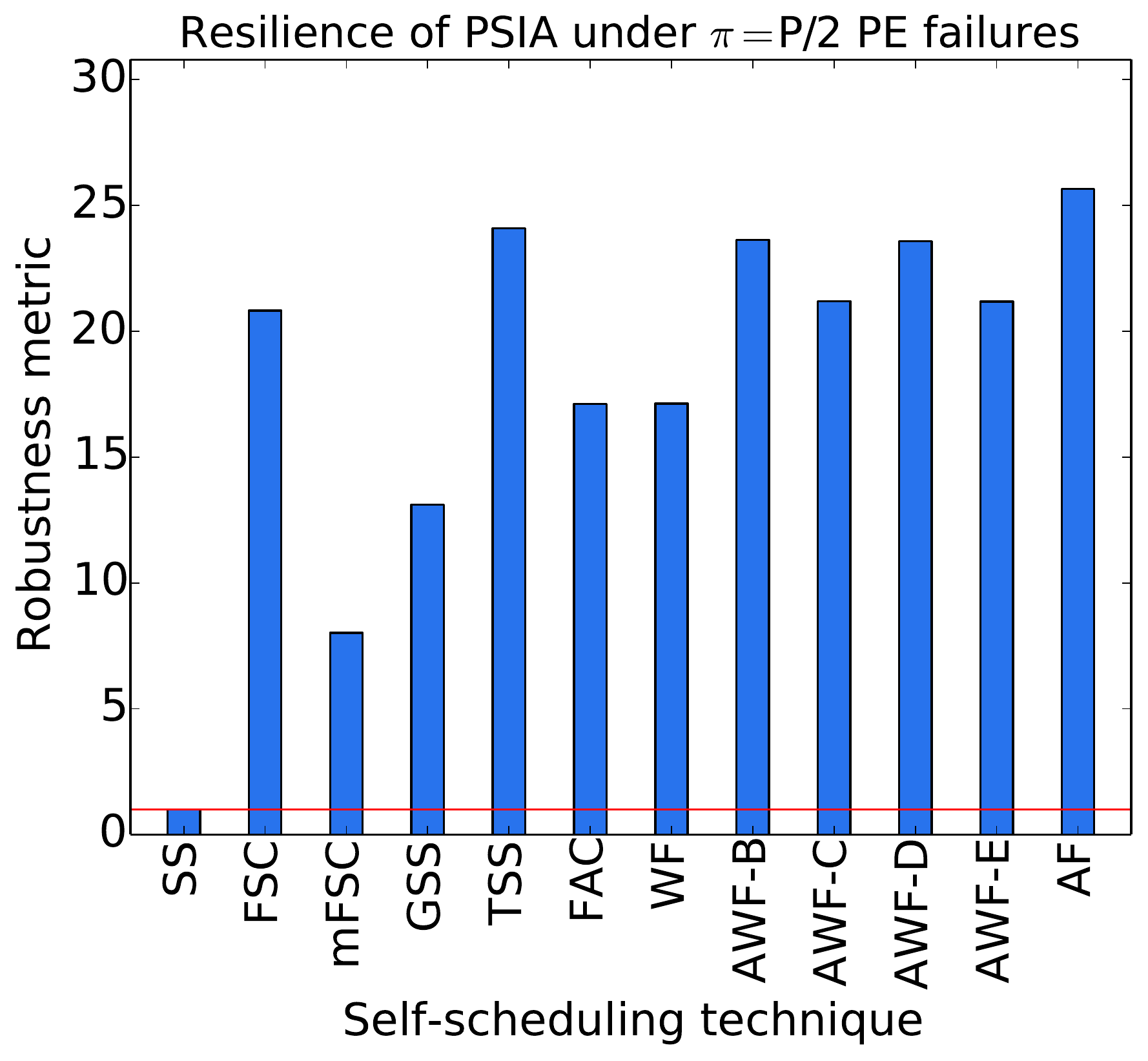}%
		\label{subfig:PSIA_ft_metric_p2}%
	} \hspace{0cm} 
	\subfloat{%
	\includegraphics[clip, trim=0cm 0cm 0cm 0cm, scale=0.24]{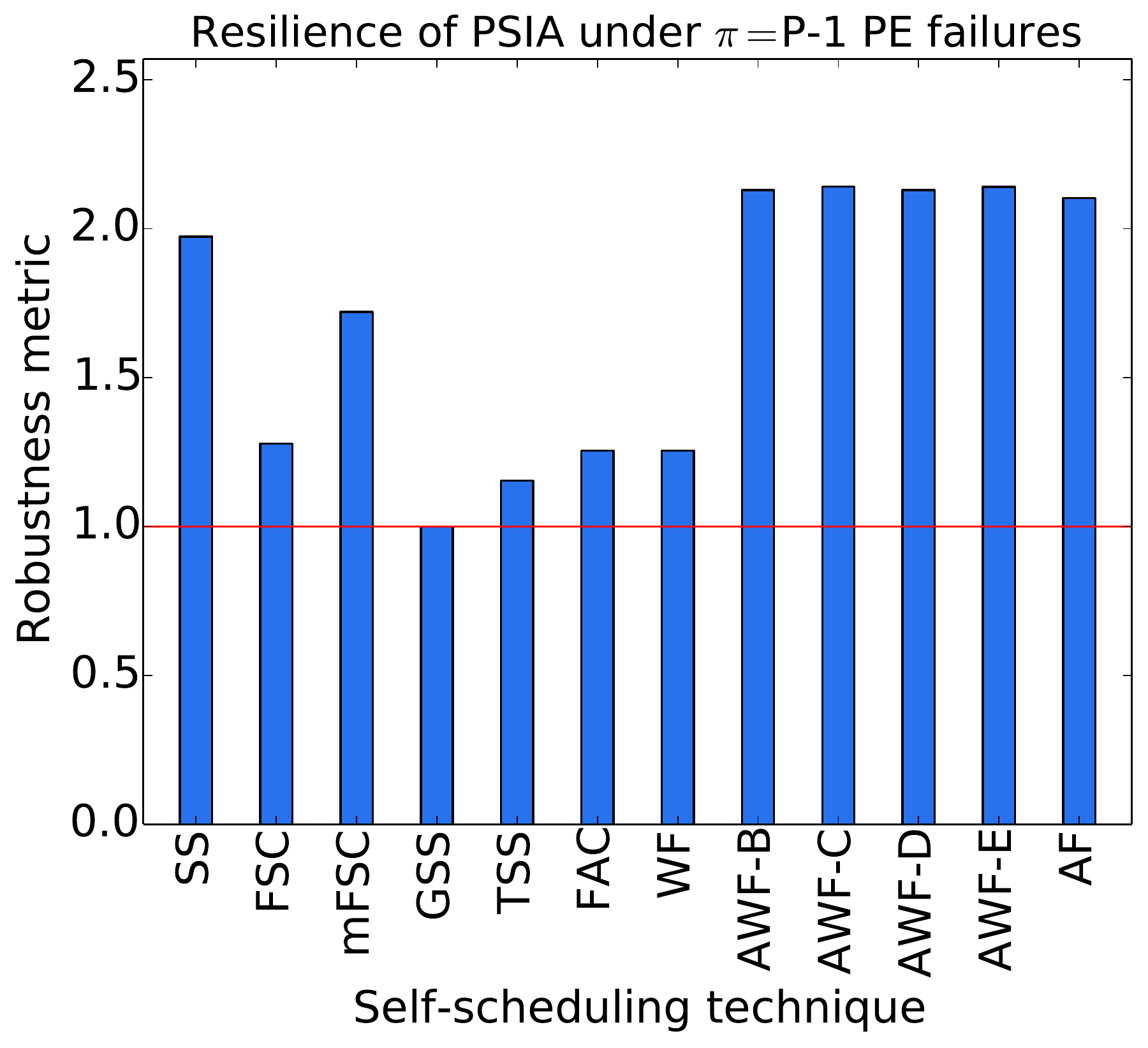}%
	\label{subfig:PSIA_ft_metric_pm1}%
} 	\\ \text{(a) PSIA - failures}\\
		\subfloat{%
			\includegraphics[clip, trim=0cm 0cm 0cm 0cm, scale=0.24]{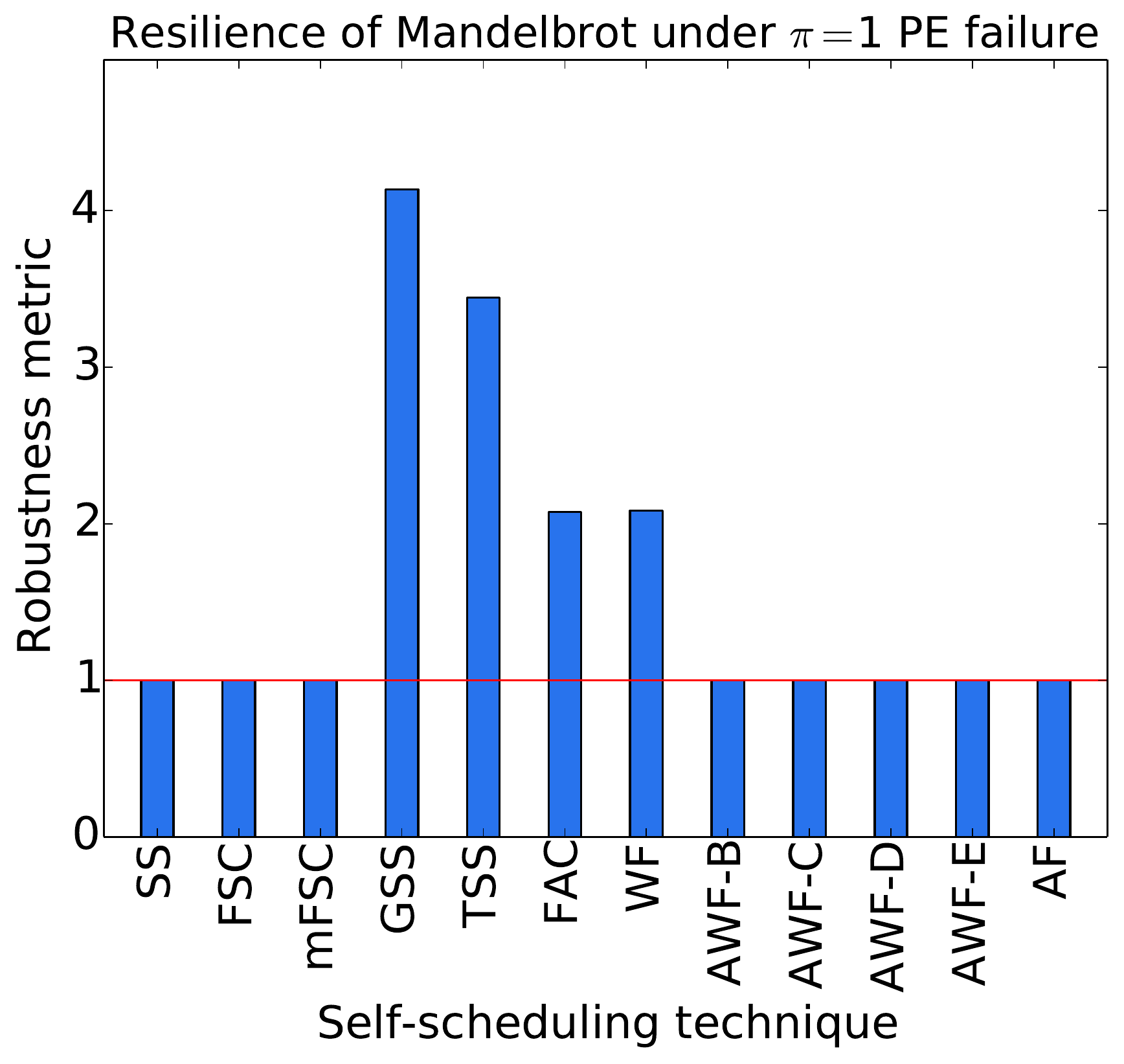}%
			\label{subfig:Mandel_ft_metric_1}%
		} \hspace{0cm}
		\subfloat{%
		\includegraphics[clip, trim=0cm 0cm 0cm 0cm, scale=0.24]{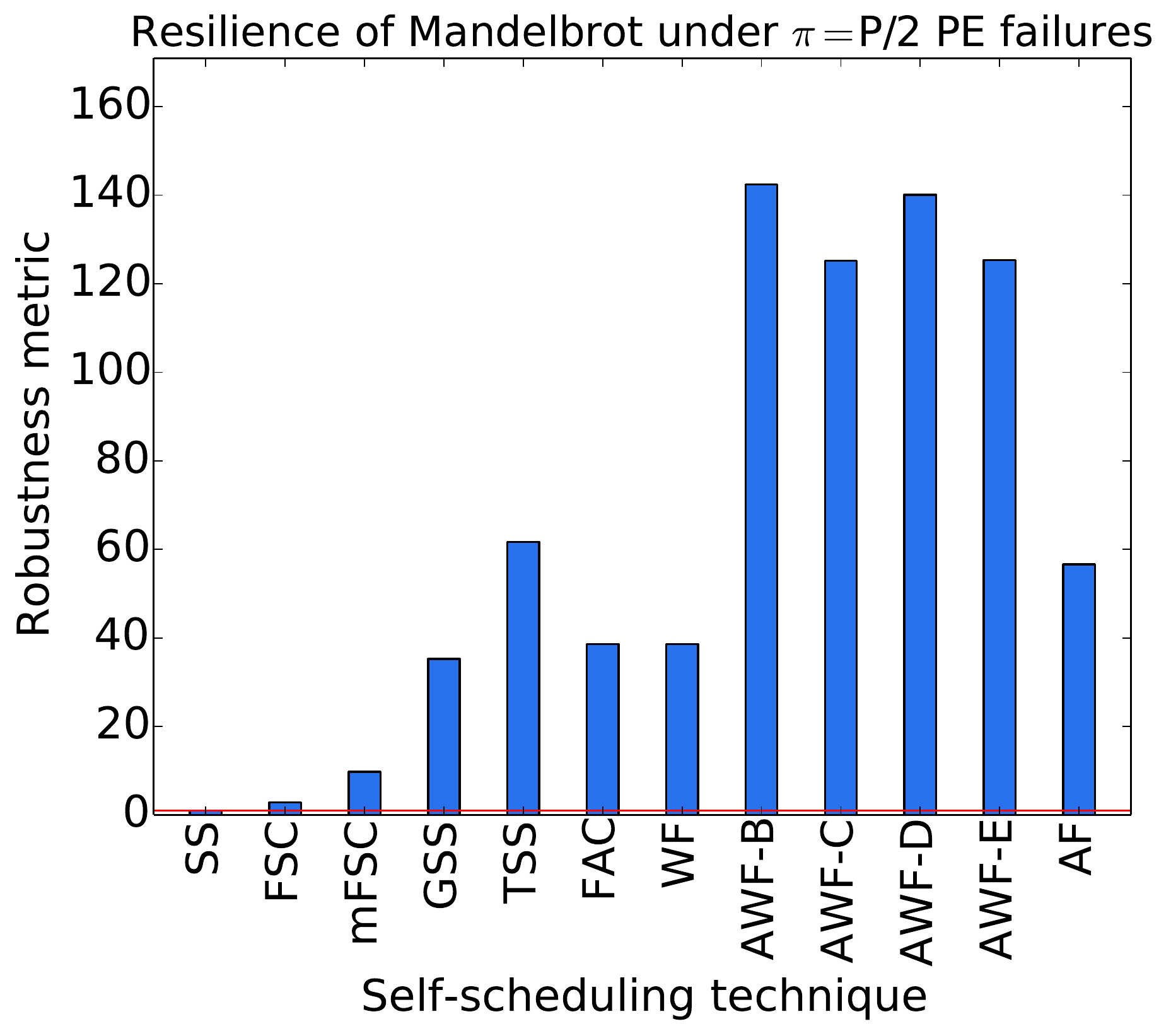}%
		\label{subfig:Mandel_ft_metric_p2}%
	}\hspace{0cm} 
	\subfloat{%
	\includegraphics[clip, trim=0cm 0cm 0cm 0cm, scale=0.24]{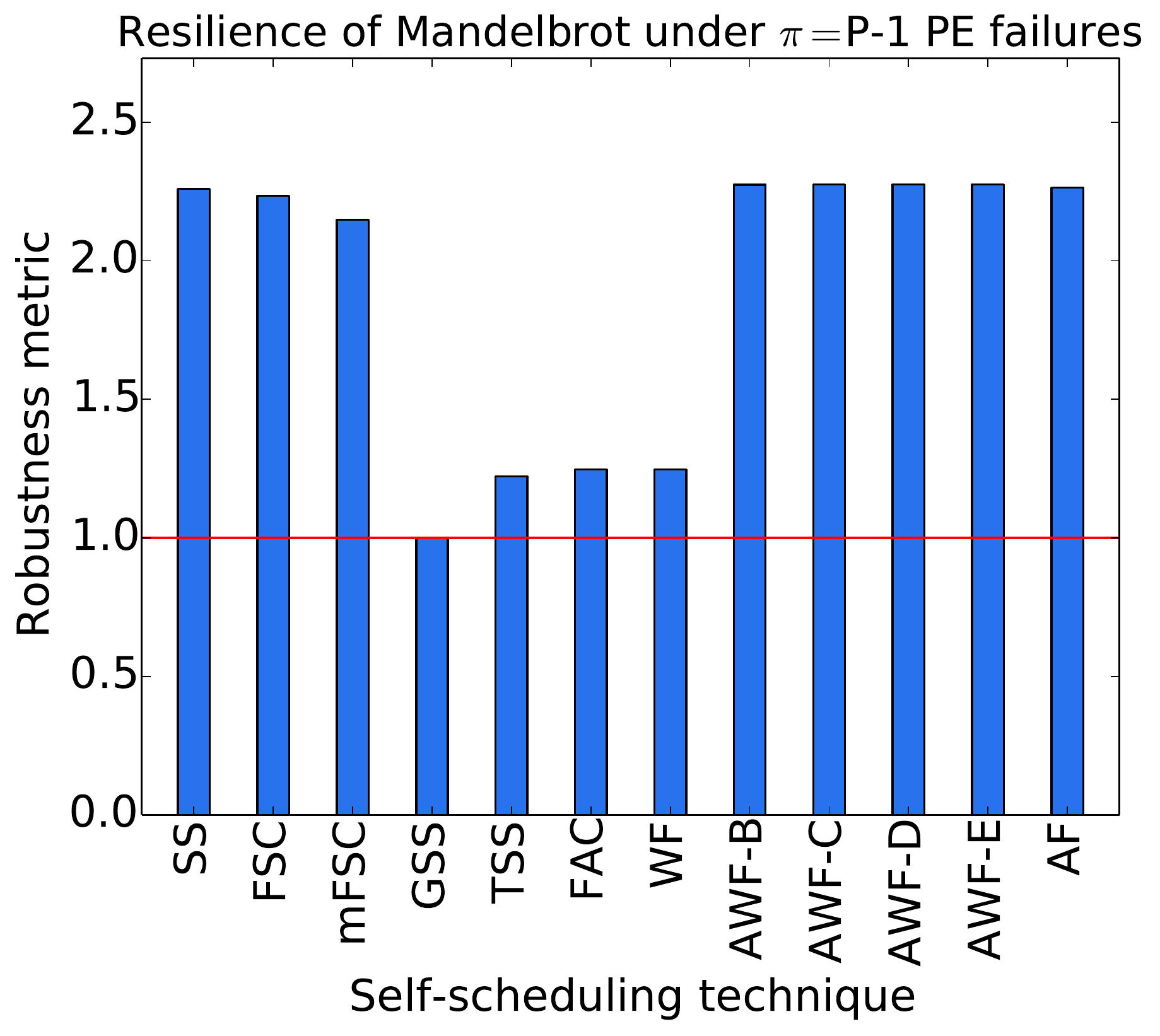}%
	\label{subfig:Mandel_ft_metric_pm1}%
} 
	\\ \text{(b) Mandelbort - failures} \\
		\caption{Resilience of DLS techniques executing PSIA and Mandelbrot with the \rdlb{} under failures on miniHPC with $256$ cores. The metrics show how many folds is a DLS technique robust with respect to the most robust DLS technique (with metric = $1$) in a particular failure scenario (lower is better).}
		\label{fig:metrics_ft}
	\end{minipage}
\end{figure*}

\begin{figure*}[htbp]
	\begin{minipage}{\textwidth}
		\centering
		\subfloat{%
			\includegraphics[clip, trim=0cm 0cm 0cm 0cm, scale=0.24]{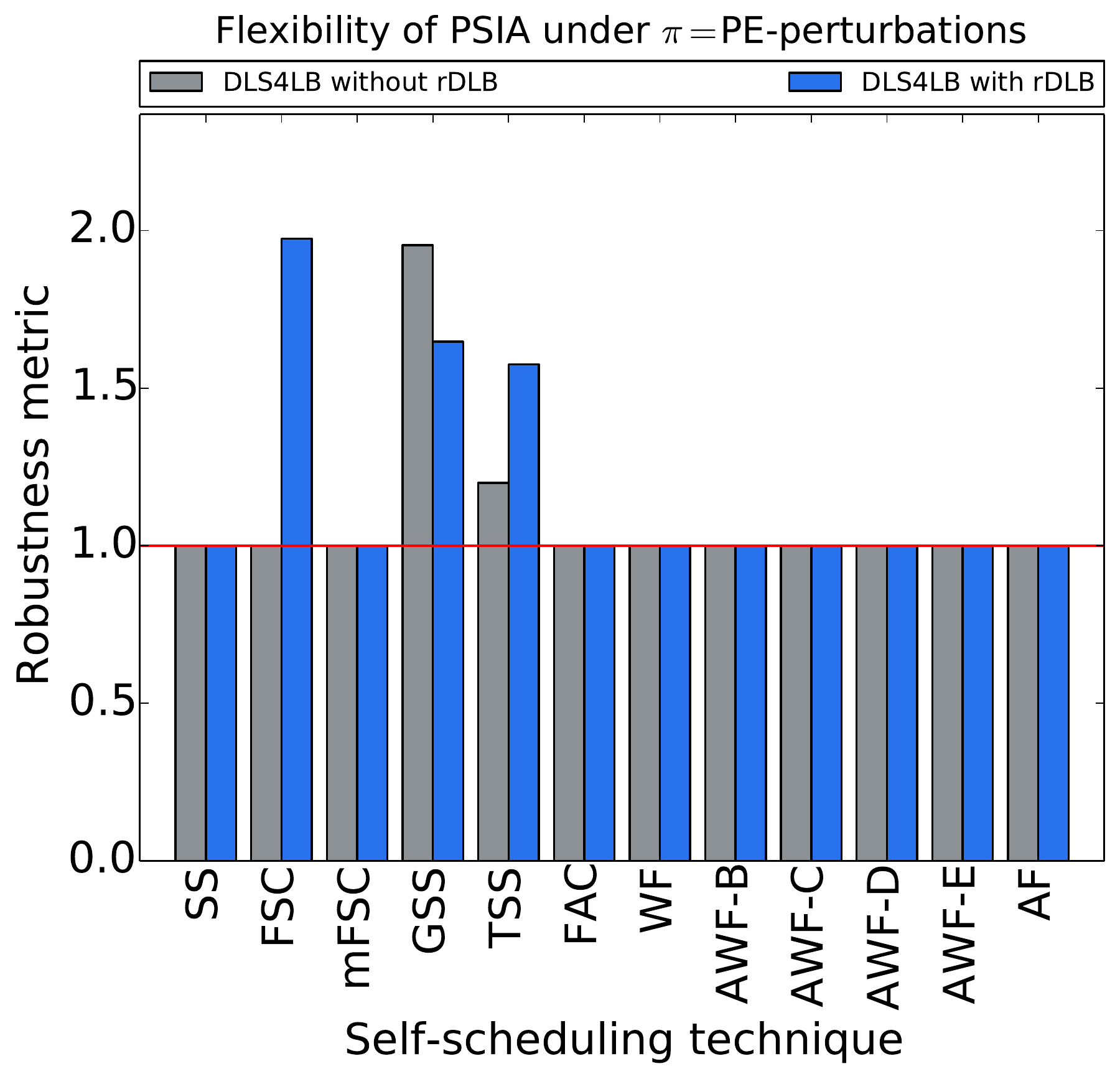}%
			\label{subfig:PSIA_pert_metric_pe}%
		} \hspace{0cm} 
		\subfloat{%
			\includegraphics[clip, trim=0cm 0cm 0cm 0cm, scale=0.24]{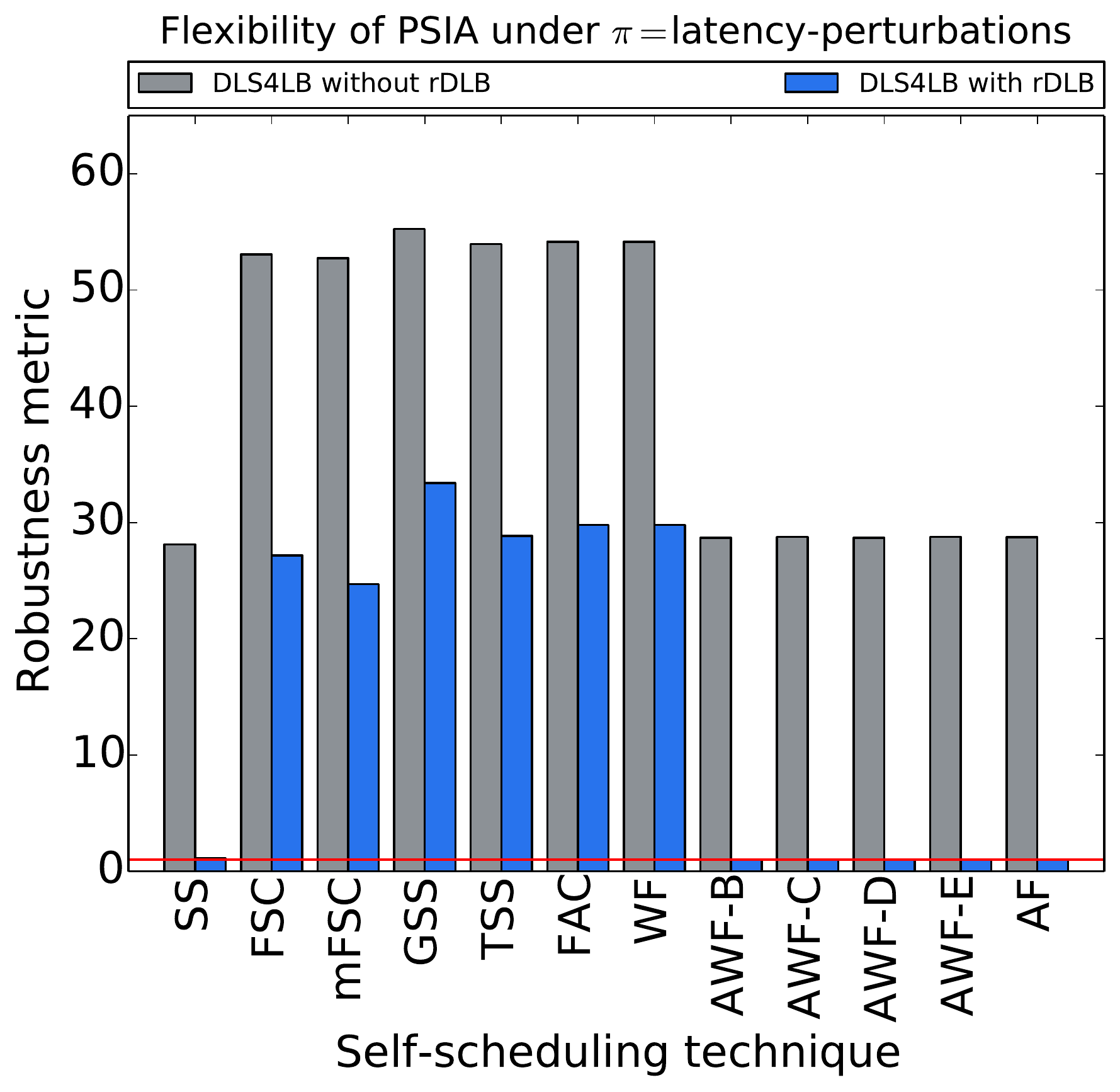}%
			\label{subfig:PSIA_pert_metric_lat}%
		} \hspace{0cm} 
		\subfloat{%
			\includegraphics[clip, trim=0cm 0cm 0cm 0cm, scale=0.24]{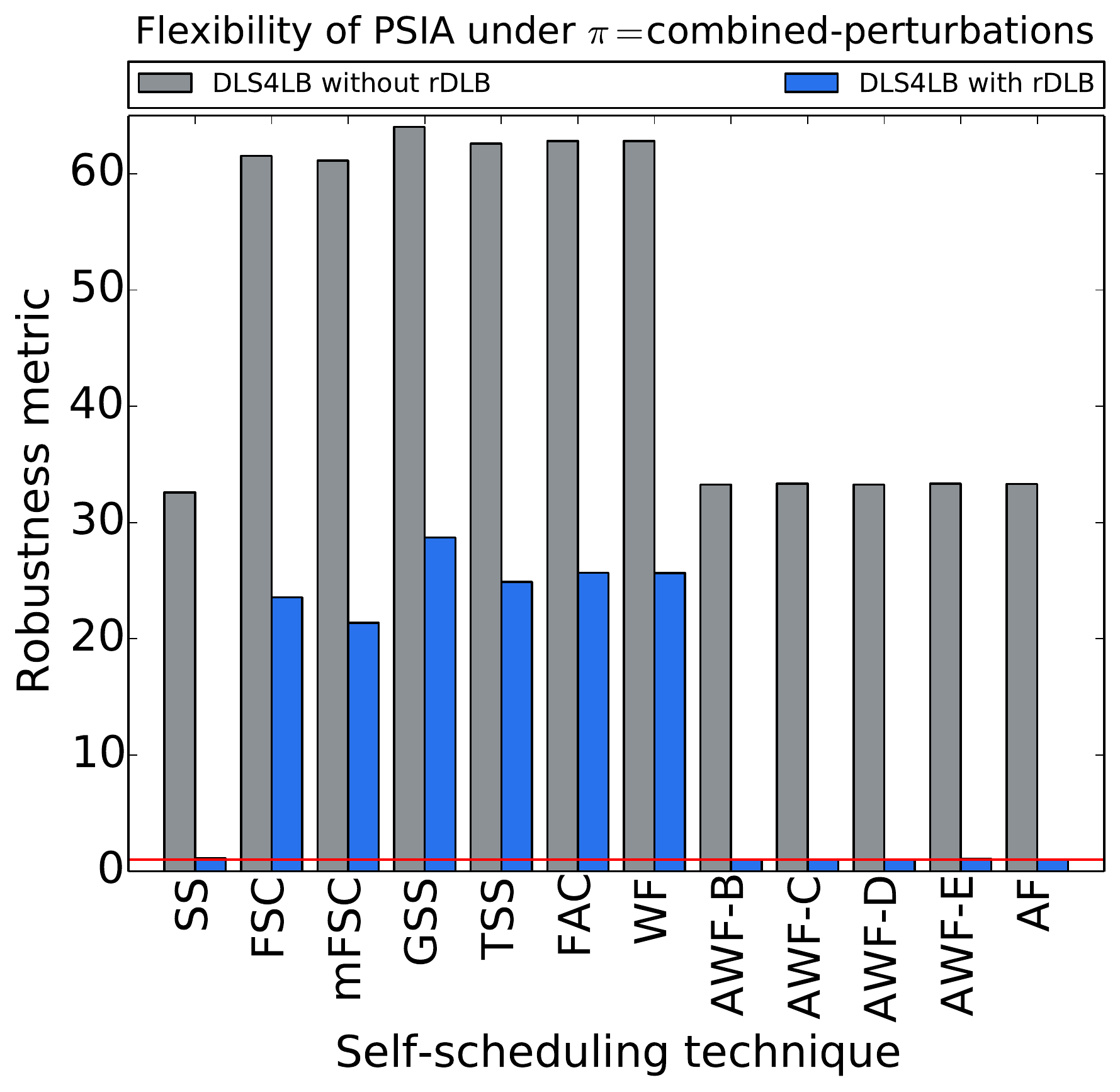}%
			\label{subfig:PSIA_pert_metric_pelat}%
		} 	\\ \text{(a) PSIA - perturbations}\\
		\subfloat{%
			\includegraphics[clip, trim=0cm 0cm 0cm 0cm, scale=0.24]{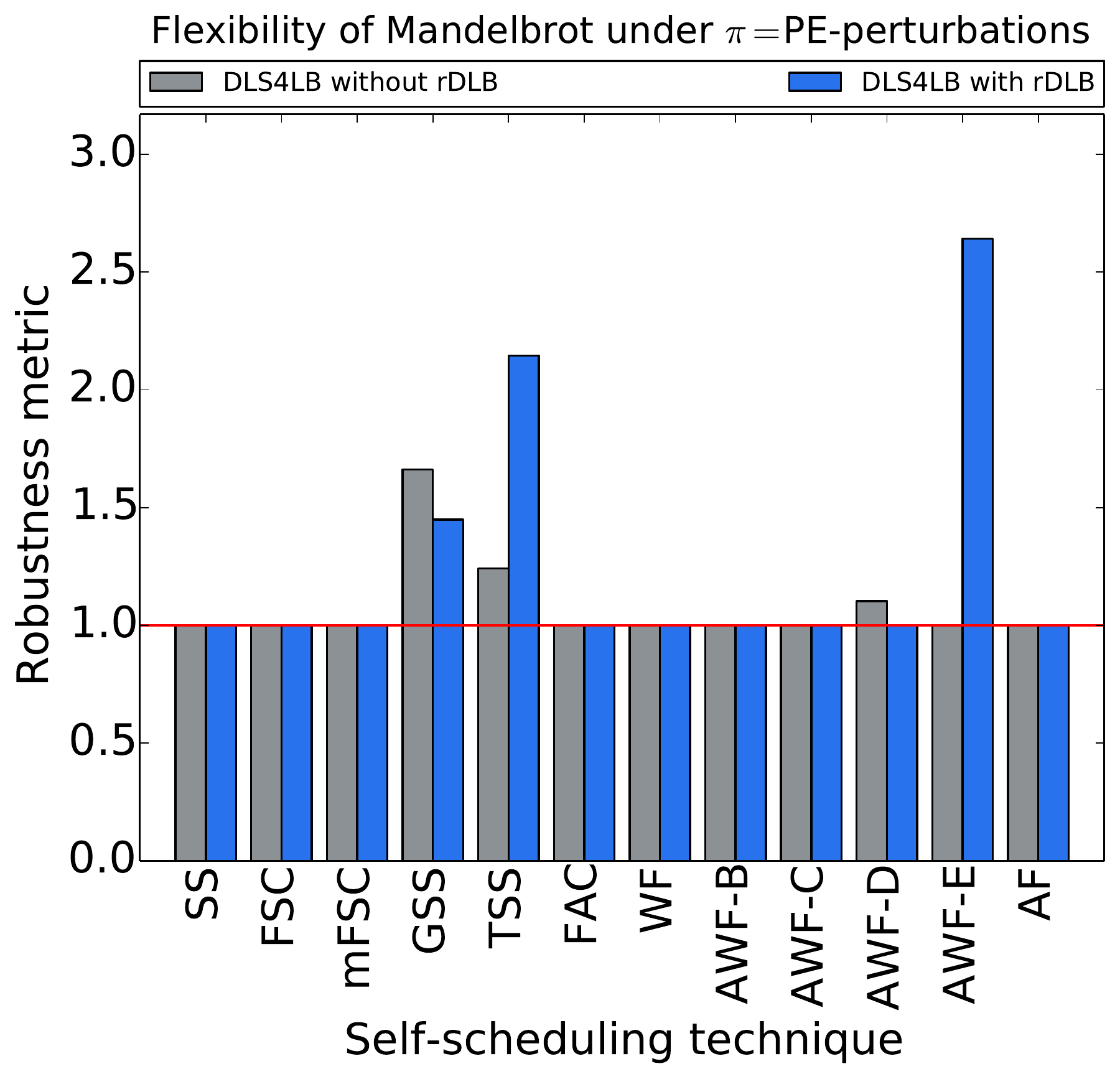}%
			\label{subfig:Mandel_pert_metric_pe}%
		} \hspace{0cm}
		\subfloat{%
			\includegraphics[clip, trim=0cm 0cm 0cm 0cm, scale=0.24]{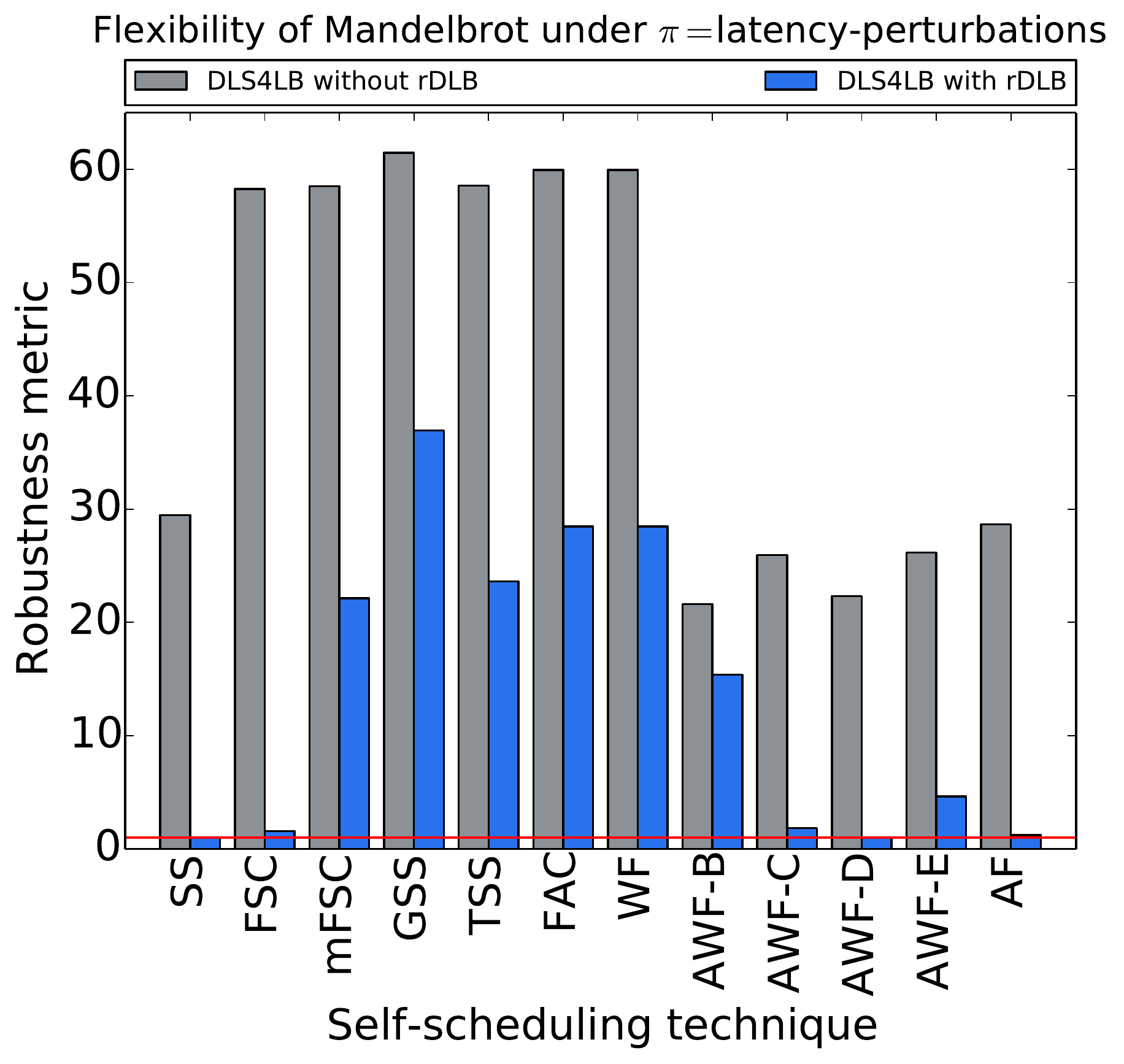}%
			\label{subfig:Mandel_pert_metric_lat}%
		}\hspace{0cm} 
		\subfloat{%
			\includegraphics[clip, trim=0cm 0cm 0cm 0cm, scale=0.24]{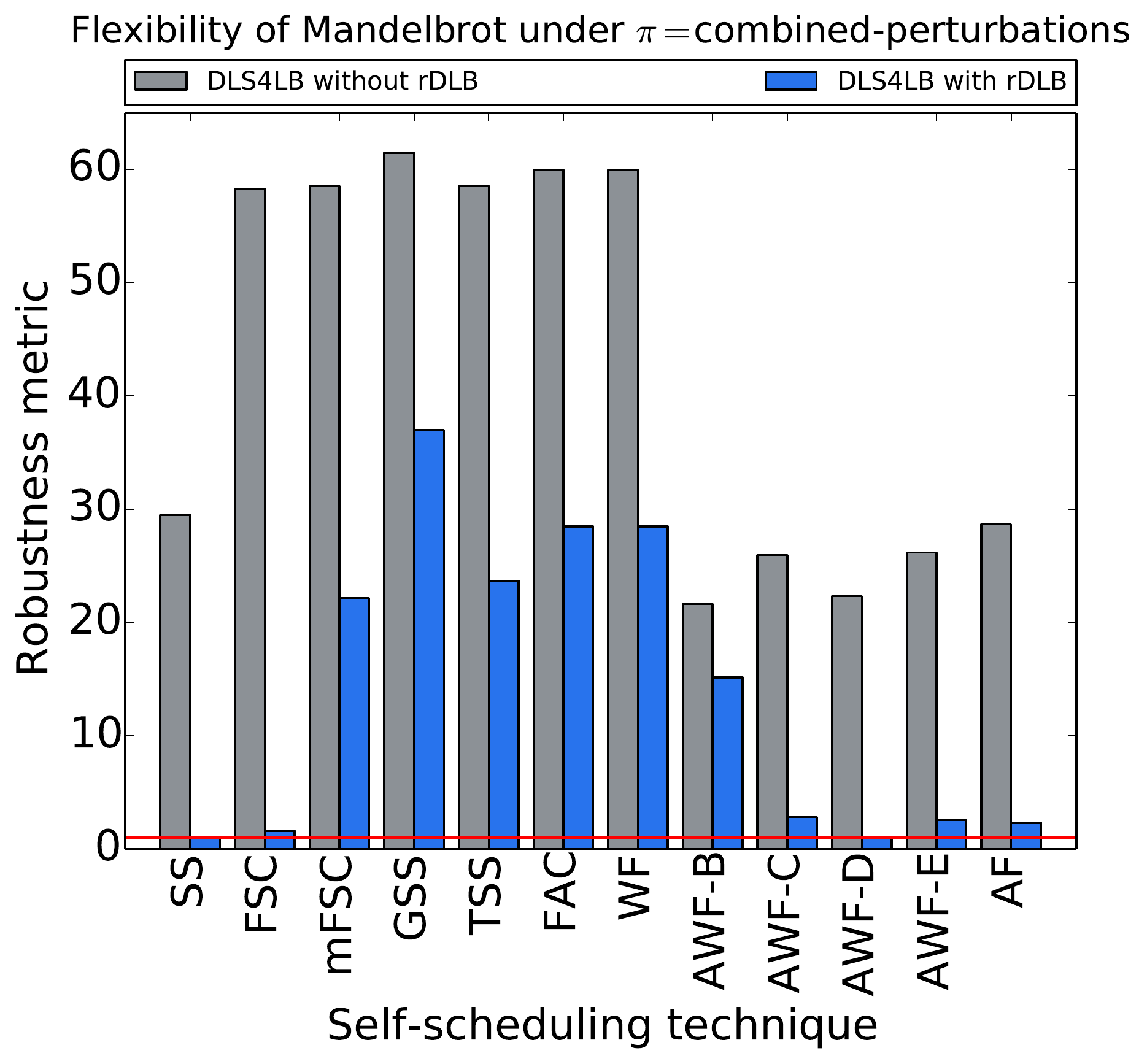}%
			\label{subfig:Mandel_pert_metric_pelat}%
		} 
		\\ \text{(b) Mandelbort - perturbations} \\
		\caption{Flexibility of DLS techniques executing PSIA and Mandelbrot without and with the \rdlb{} under PE perturbations, and latency perturbation scenarios on miniHPC with $256$ cores. The metrics show how many folds is a DLS technique robust with respect to the most robust DLS technique (with metric = $1$) in a particular perturbation scenario (lower is better).}
		\label{fig:metrics_pert}
	\end{minipage}
\end{figure*}

\subsection{Evaluation and Discussion}
The performance results of PSIA and Mandelbrot under failures and perturbations is depicted in \figurename{~\ref{fig:Tpar}} and the robustness metrics are presented in \figurename{~\ref{fig:metrics_ft}} and \figurename{~\ref{fig:metrics_pert}}.
STATIC technique is not included in the results with \rdlb{} as \rdlb{} is applied for dynamic self-scheduling techniques only.
The results show the average parallel loop execution time over $20$ executions for each experiment.
As execution with \dlbTool{} without \rdlb{} will hang forever in the presence of failures, \figurename{~\ref{subfig:PSIA_ft}}, \figurename{~\ref{subfig:Mandel_ft}}, only show the results with \rdlb{}.
Inspecting \figurename{~\ref{subfig:PSIA_ft}}, \figurename{~\ref{subfig:Mandel_ft}}, and \figurename{~\ref{fig:metrics_ft}} reveals that one PE failure is tolerated with \rdlb{} with almost no effect on the execution time, i.e., execution time with one failure is very close to the baseline.
Specifically for the adaptive DLS techniques, the execution times in the case of single failure and baseline are almost equal.
The cost of tolerating $P/2$ failures depends largely on the DLS technique.
DLS techniques that assign small chunk sizes, such as SS (the most robust in this scenario), are more robust that techniques that assign large chunks. 
Small chunk sizes in such case minimizes the amount of lost work in the case of PE failures.
In the case of $P-1$ failures, The work is almost serialized on only the master, and in this case the execution time depends on overhead of the DLS technique (number of scheduling rounds).
For the execution with perturbations, two experiments are performed per scheduling technique per perturbation scenario: without \rdlb{} and with \rdlb{} to show the benefit of \rdlb{} in enhancing application performance and robustness under perturbations.
The results show that perturbations in PE availability do not affect the performance significantly. 
Comparing the performance results in latency and combined perturbations, one can see that \rdlb{} enhanced the performance of PSIA in \figurename{~\ref{subfig:PSIA_pert}} and Mandelbrot in \figurename{~\ref{subfig:Mandel_pert}}. 
Results of the flexibility metric in \figurename{~\ref{fig:metrics_pert}} confirm that using \rdlb{} approach boosted the robustness of DLS techniques in case of severe latency perturbations and combined PE and latency availabilities perturbations.
In fact,  for the adaptive techniques AWF-B,-C, -D, -E their flexibility is enhanced more than $30$ folds by applying \rdlb{} approach in the case of combined PE and latency perturbations in PSIA execution.

\begin{figure*}[htbp]
	\begin{minipage}{\textwidth}
		\centering
		\subfloat[PSIA - Baseline]{%
			\includegraphics[clip, trim=0cm 0cm 0cm 0cm, scale=0.24]{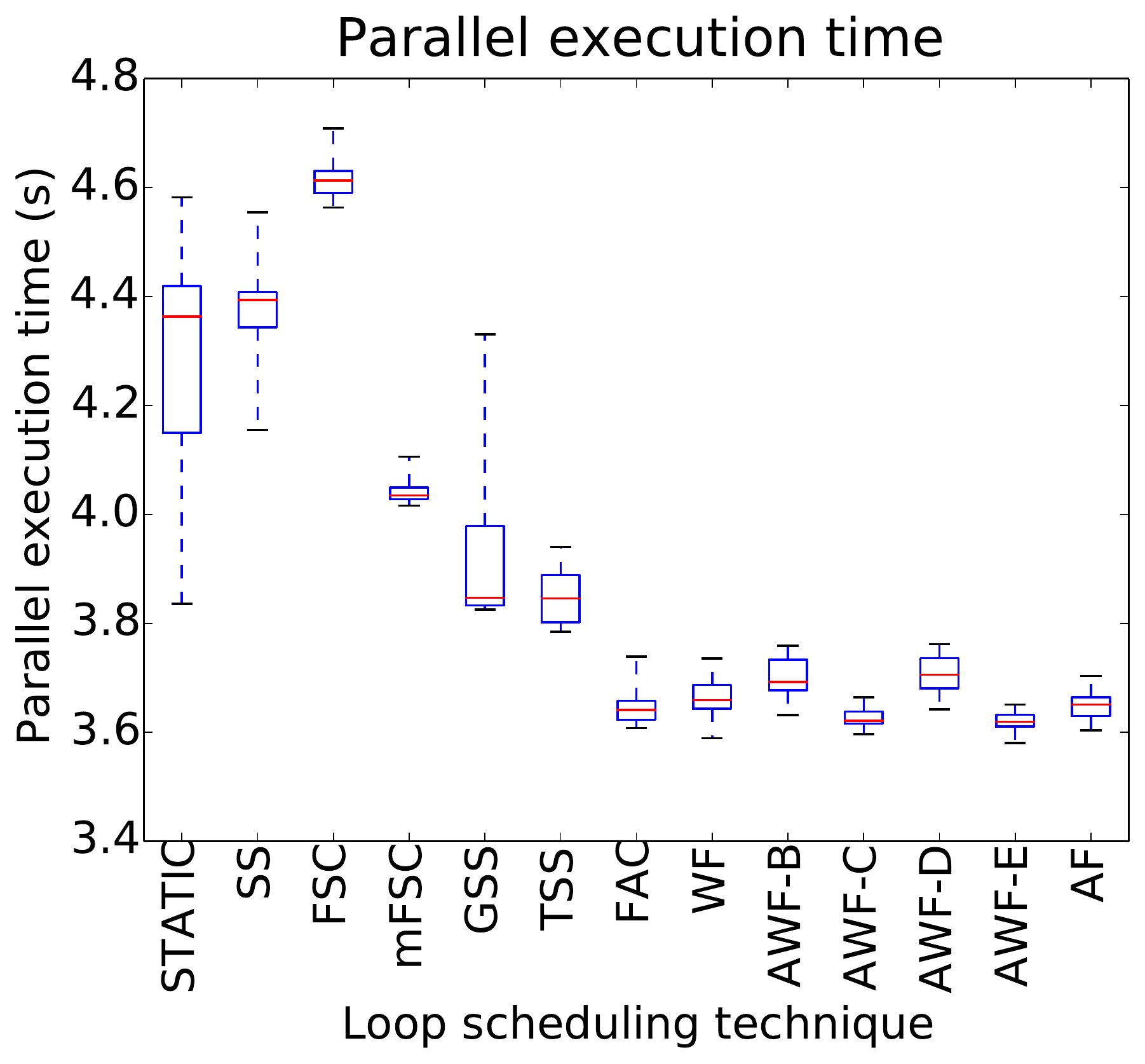}%
			\label{subfig:PSIA_baselinet}%
		} \hspace{0cm} 
		\subfloat[Mandelbrot - Baseline]{%
			\includegraphics[clip, trim=0cm 0cm 0cm 0cm, scale=0.24]{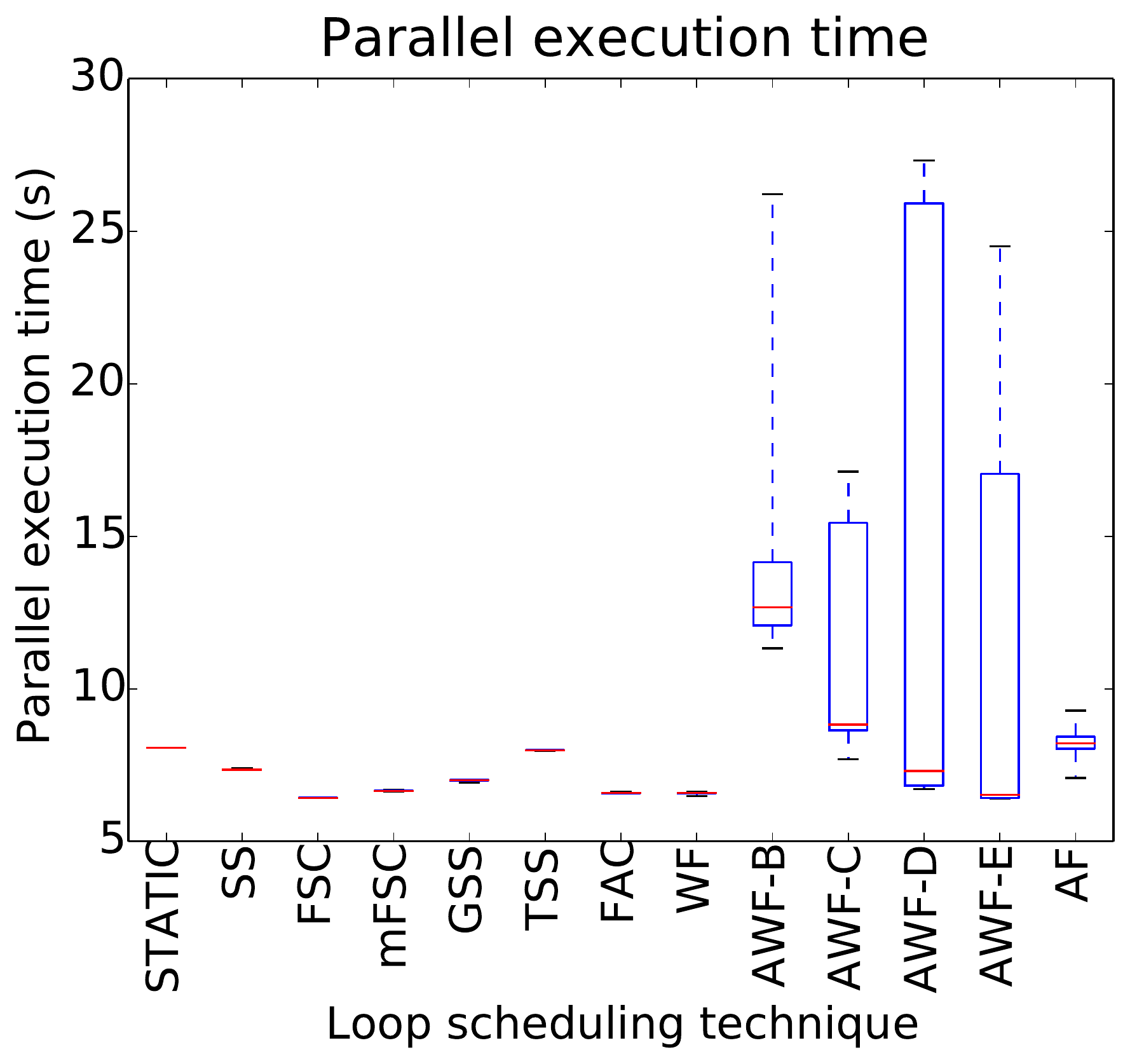}%
			\label{subfig:Mandel_baseline}%
		} \\
		\subfloat[PSIA - One failure]{%
		\includegraphics[clip, trim=0cm 0cm 0cm 0cm, scale=0.24]{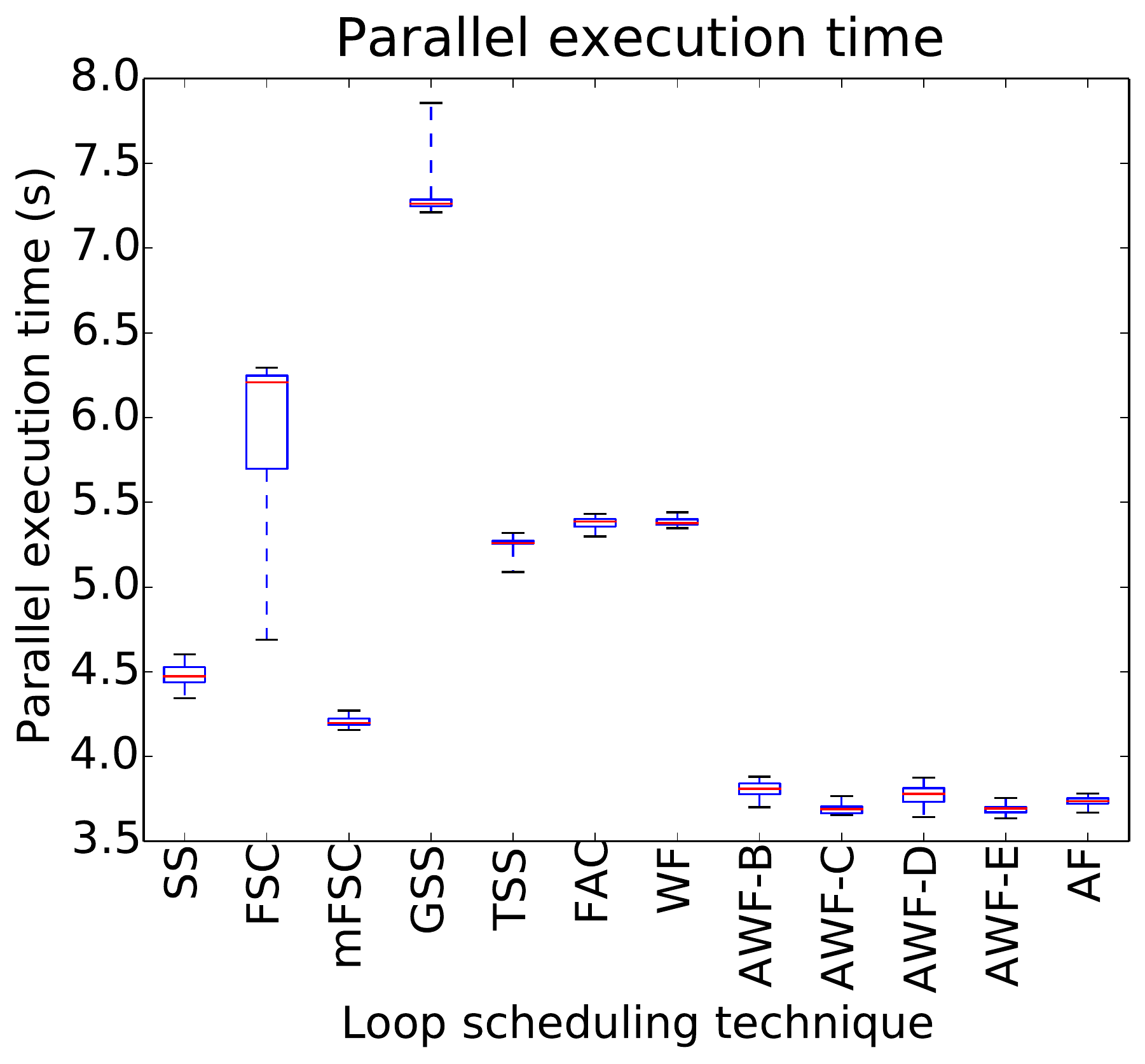}%
		\label{subfig:PSIA_1F}%
	} \hspace{0cm} 
	\subfloat[Mandelbrot - One failure]{%
		\includegraphics[clip, trim=0cm 0cm 0cm 0cm, scale=0.24]{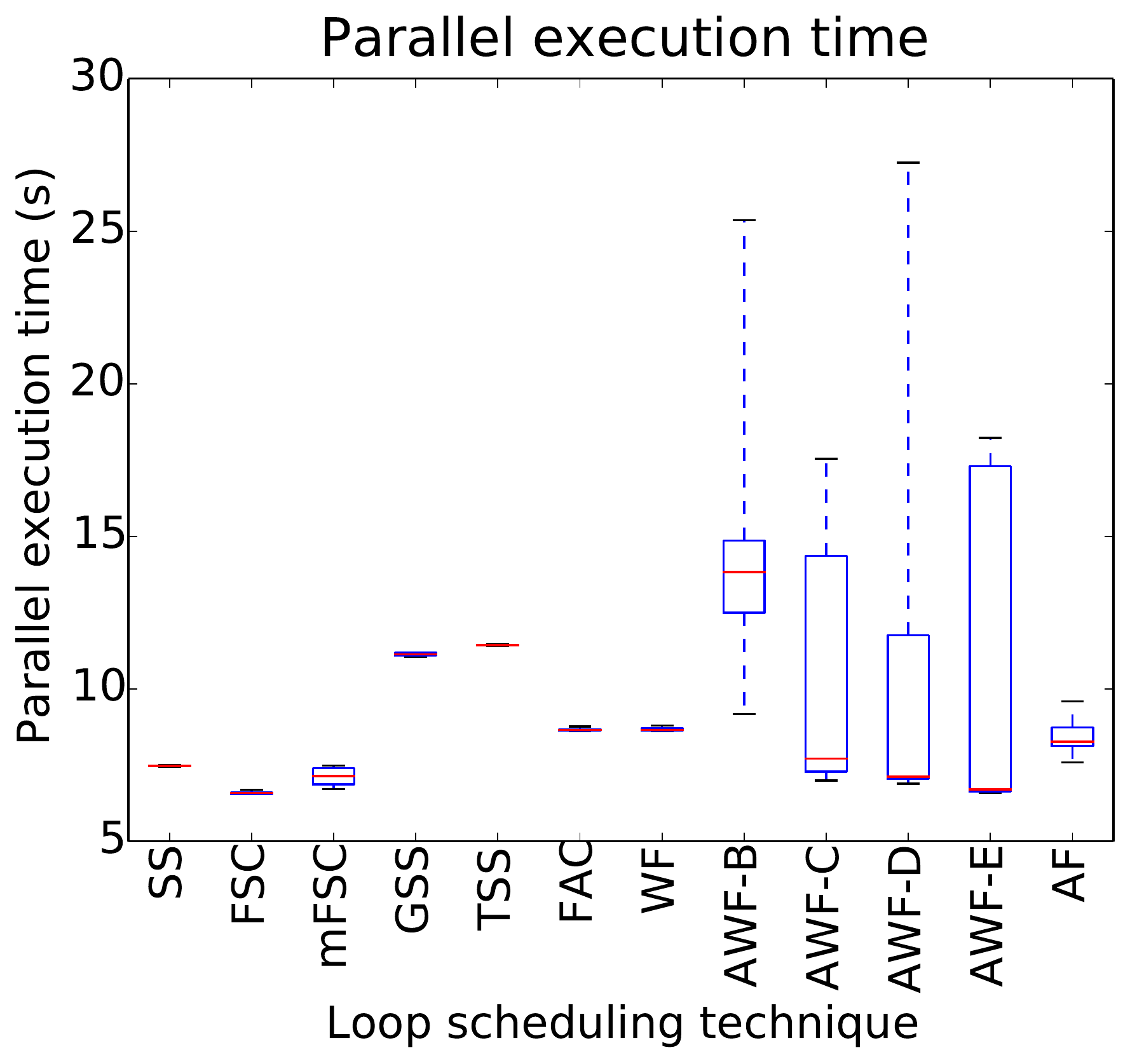}%
		\label{subfig:Mandel_1F}%
	} \\	
	\subfloat[PSIA - $P/2$ failures]{%
	\includegraphics[clip, trim=0cm 0cm 0cm 0cm, scale=0.24]{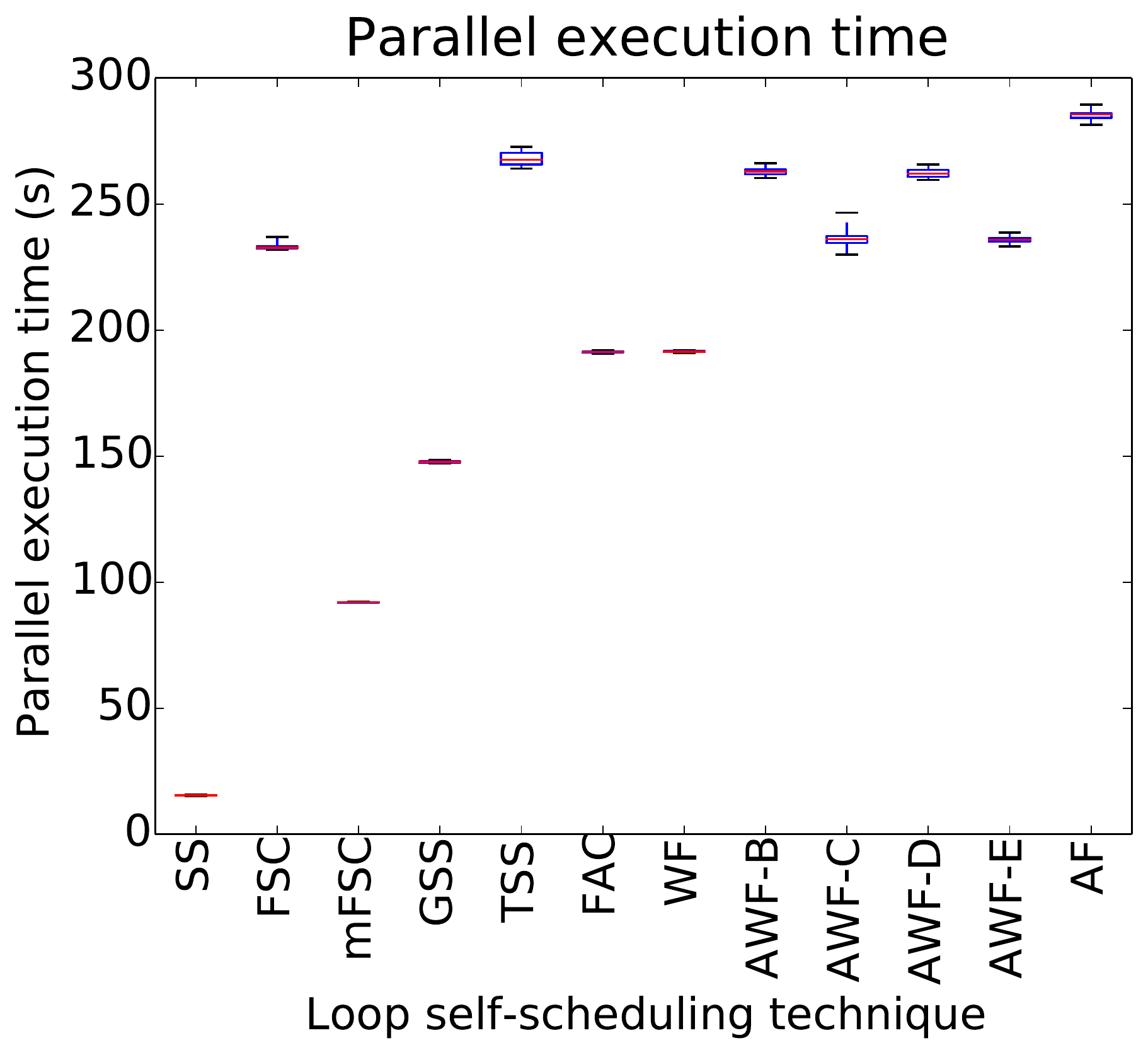}%
	\label{subfig:PSIA_P_2F}%
} \hspace{0cm} 
\subfloat[Mandelbrot -  $P/2$ failures]{%
	\includegraphics[clip, trim=0cm 0cm 0cm 0cm, scale=0.24]{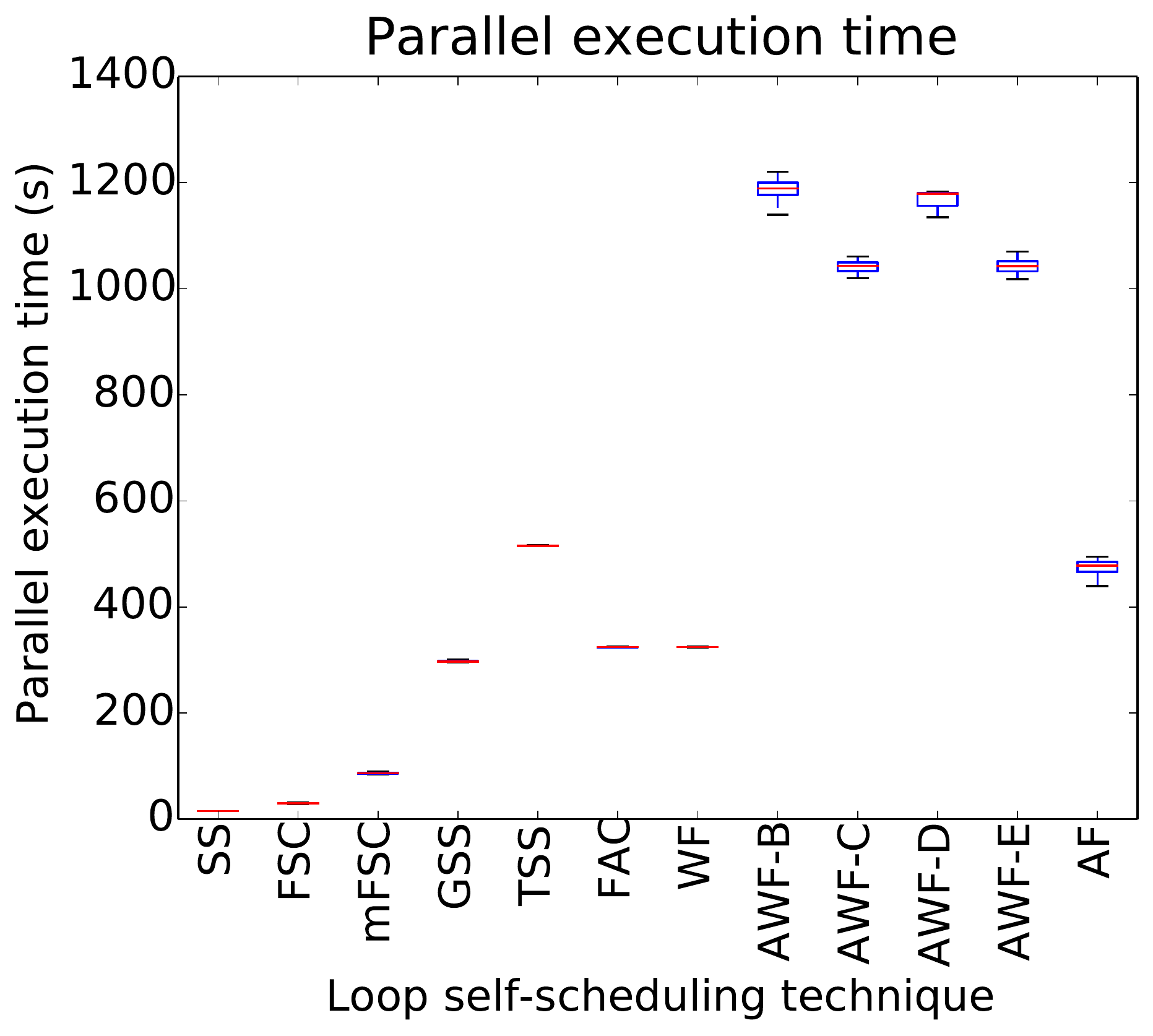}%
	\label{subfig:Mandel_P_2F}%
} \\	
	\subfloat[PSIA -  $P-1$ failures]{%
	\includegraphics[clip, trim=0cm 0cm 0cm 0cm, scale=0.24]{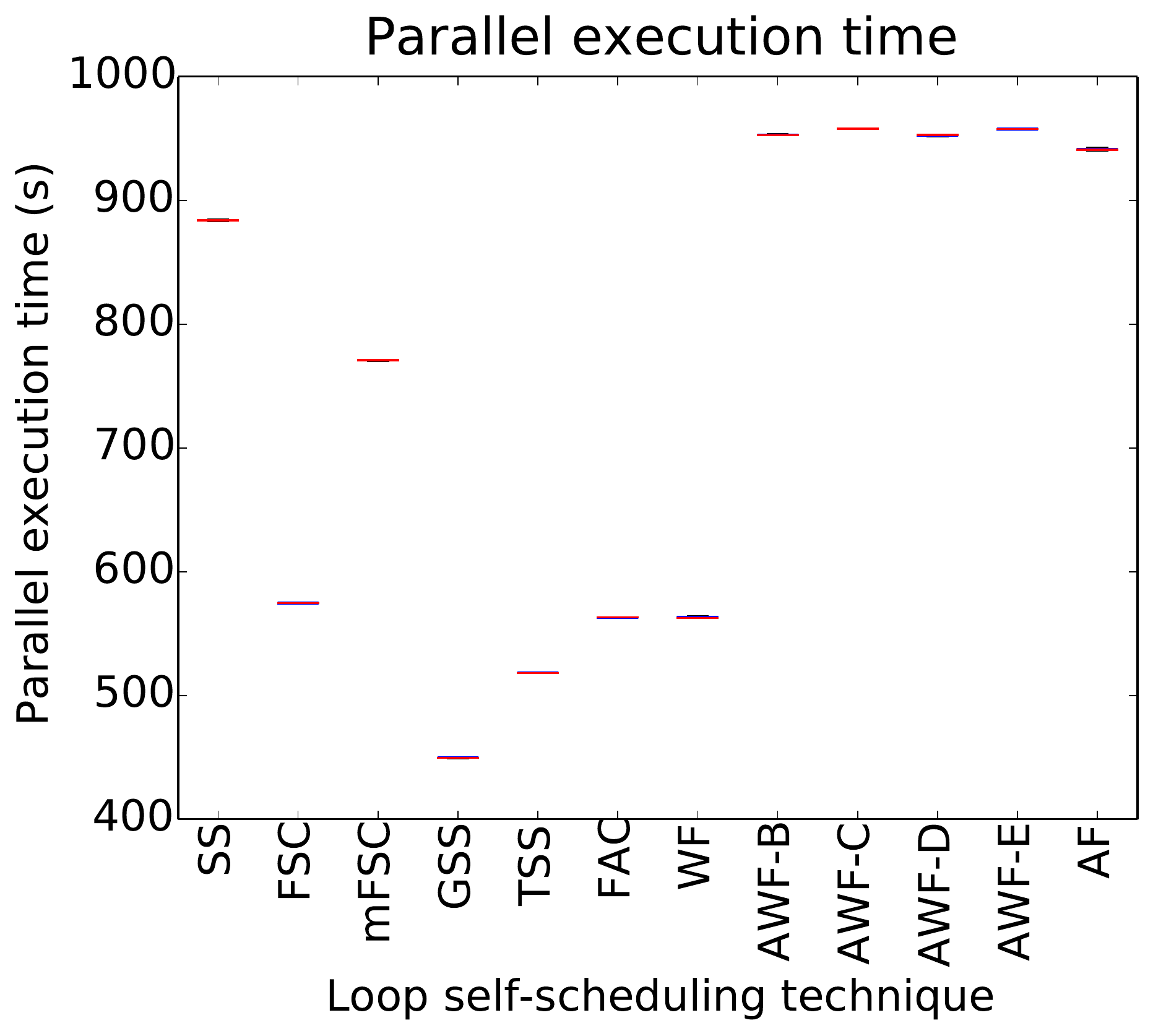}%
	\label{subfig:PSIA_P_m1F}%
} \hspace{0cm} 
\subfloat[Mandelbrot -  $P-1$ failures]{%
	\includegraphics[clip, trim=0cm 0cm 0cm 0cm, scale=0.24]{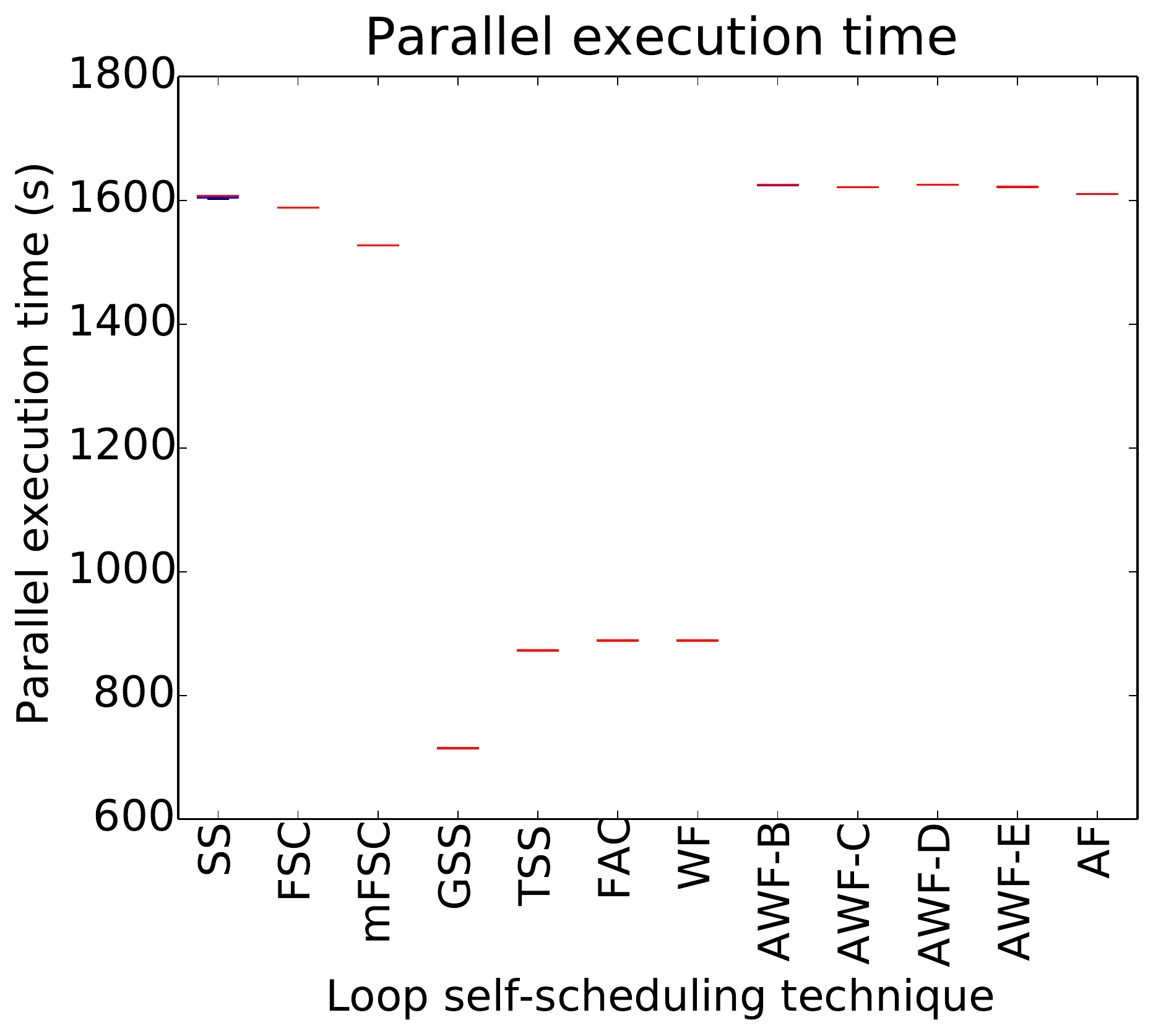}%
	\label{subfig:Mandel_P_m1F}%
} \\		
		\caption{Performance with robust \dlbTool{} under failures.}
		\label{fig:failures}
	\end{minipage}
\end{figure*}

\begin{figure*}[htbp]
	\begin{minipage}{\textwidth}
		\centering
		\subfloat[PSIA - PE perturbations, non robust]{%
			\includegraphics[clip, trim=0cm 0cm 0cm 0cm, scale=0.24]{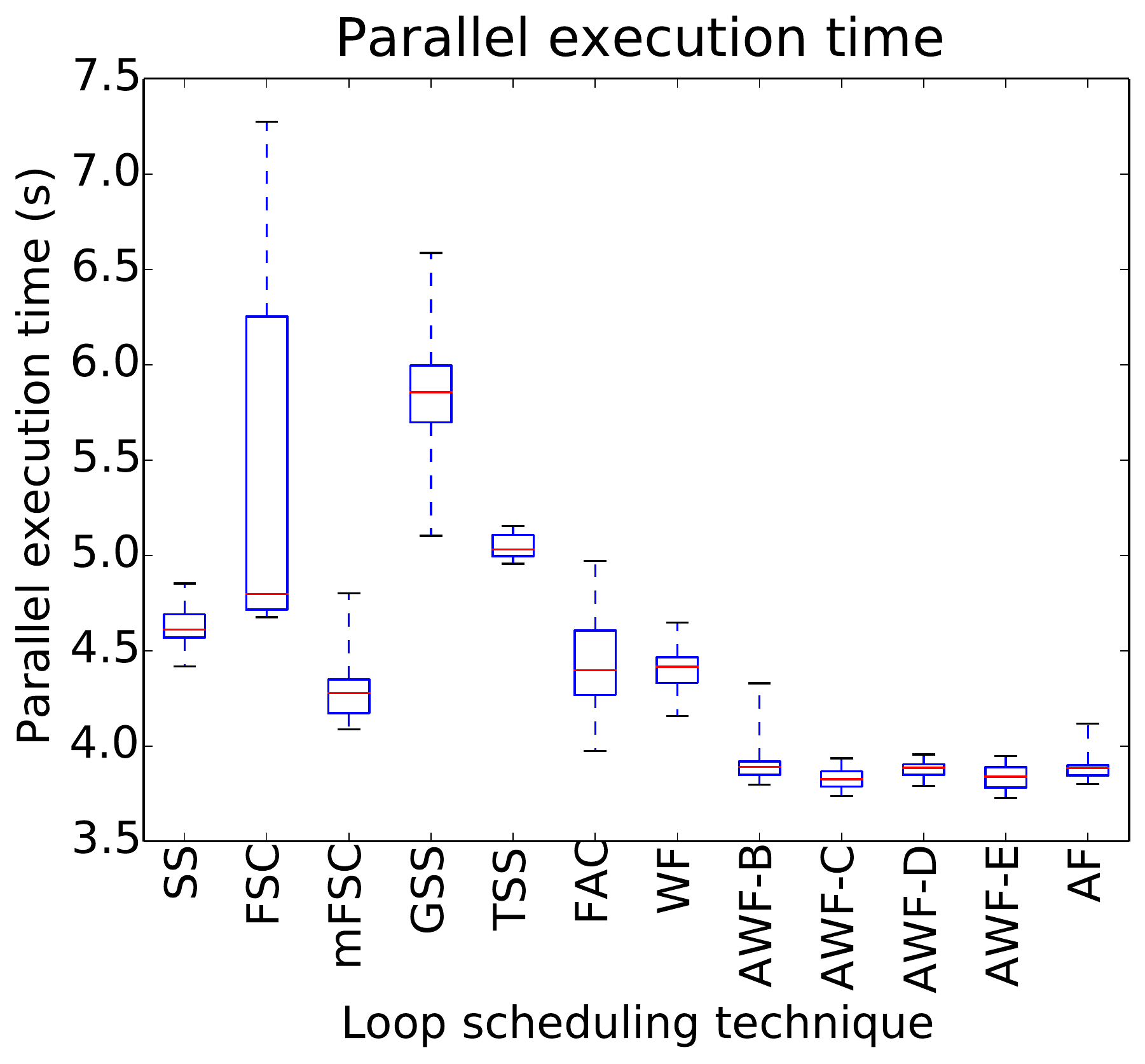}%
			\label{subfig:PSIA_PE_NOFT}%
		} \hspace{0cm} 
		\subfloat[PSIA - Network latency perturbations, non robust]{%
		\includegraphics[clip, trim=0cm 0cm 0cm 0cm, scale=0.24]{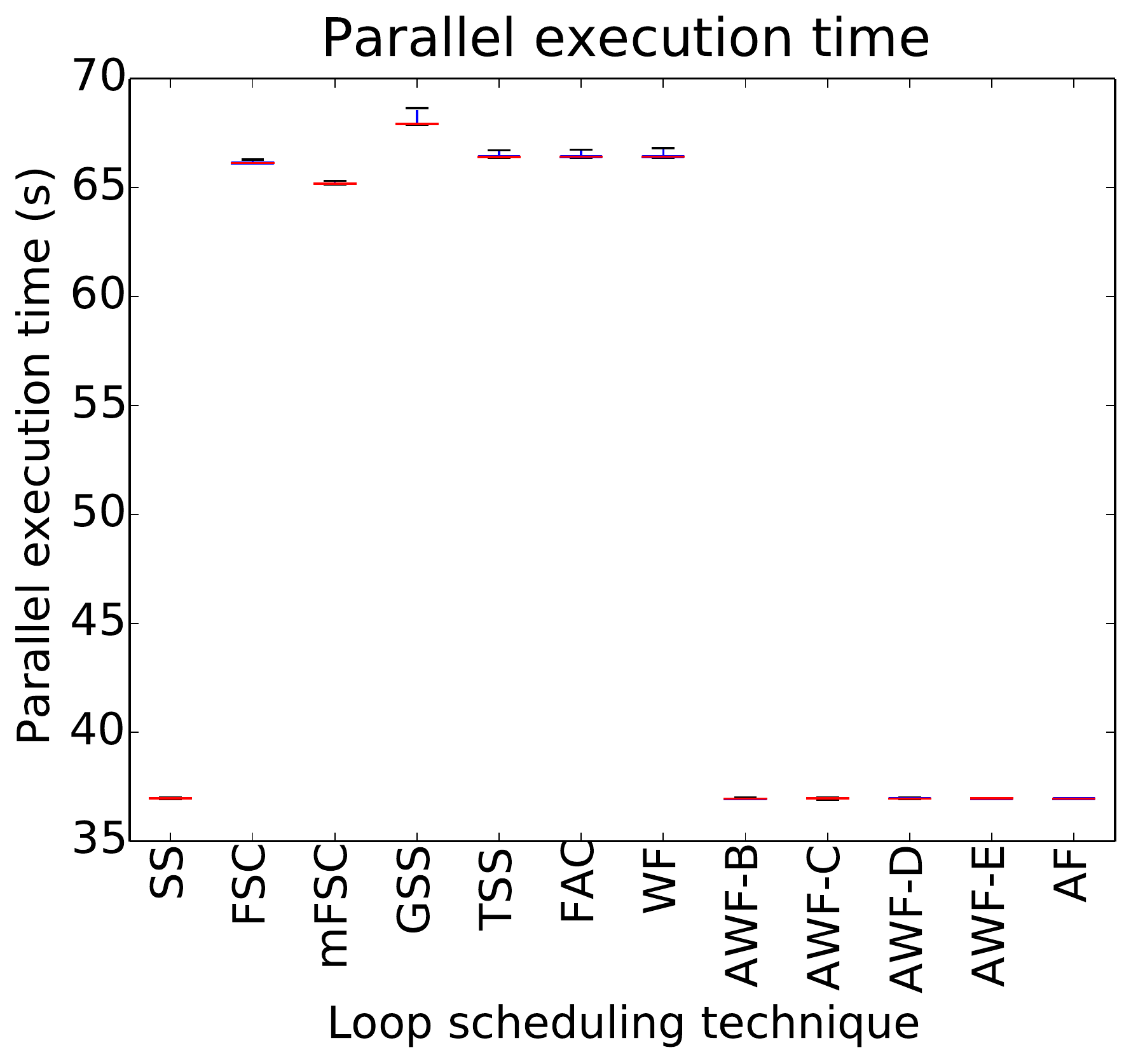}%
		\label{subfig:PSIA_lat_NOFT} 
	    }\hspace{0cm} 
		\subfloat[PSIA - Combined PE and latency perturbations, non robust]{%
		\includegraphics[clip, trim=0cm 0cm 0cm 0cm, scale=0.24]{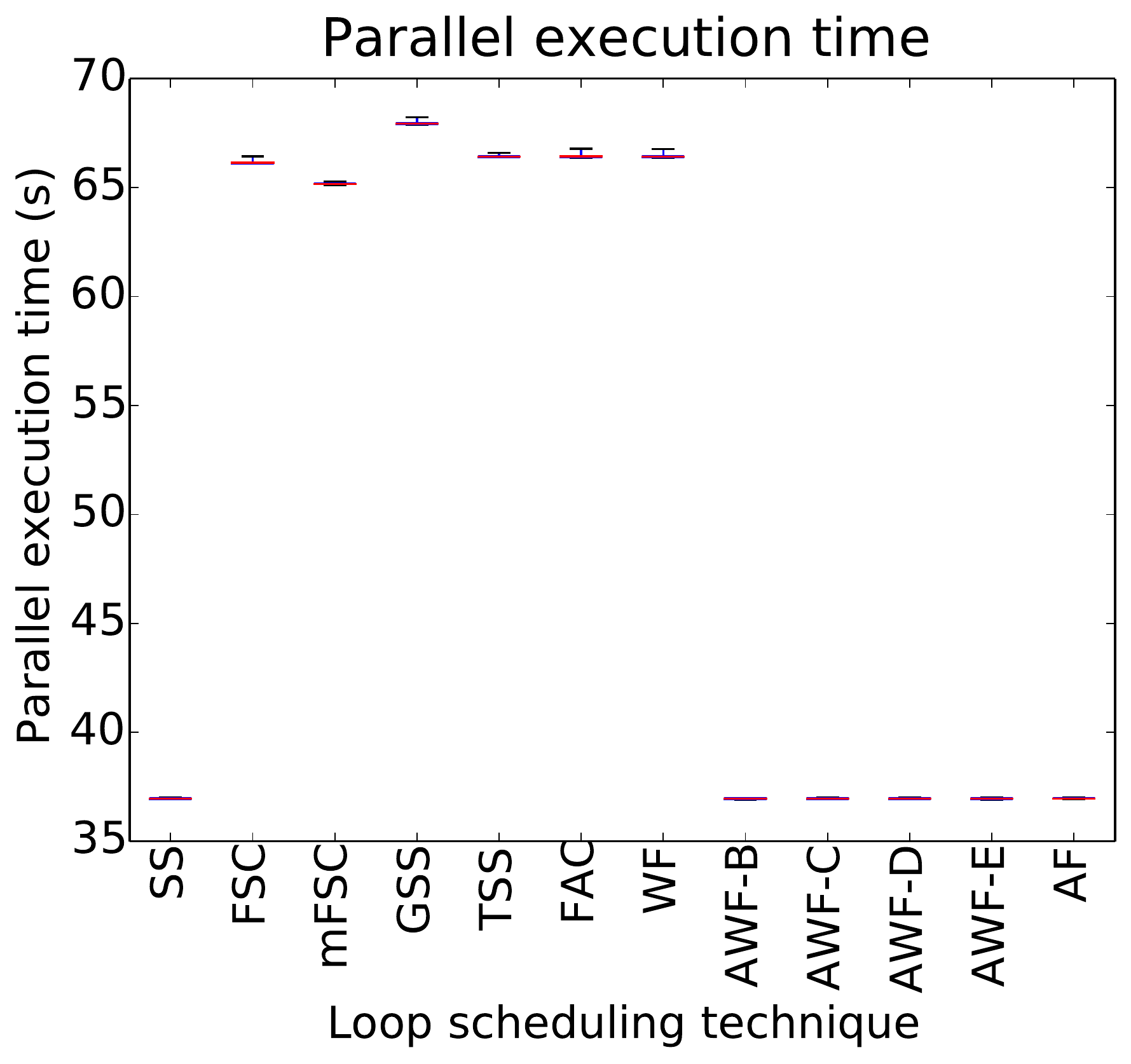}%
		\label{subfig:PSIA_pelat_NOFT}
	    } \\
		\subfloat[PSIA - PE perturbations, robust]{%
			\includegraphics[clip, trim=0cm 0cm 0cm 0cm, scale=0.24]{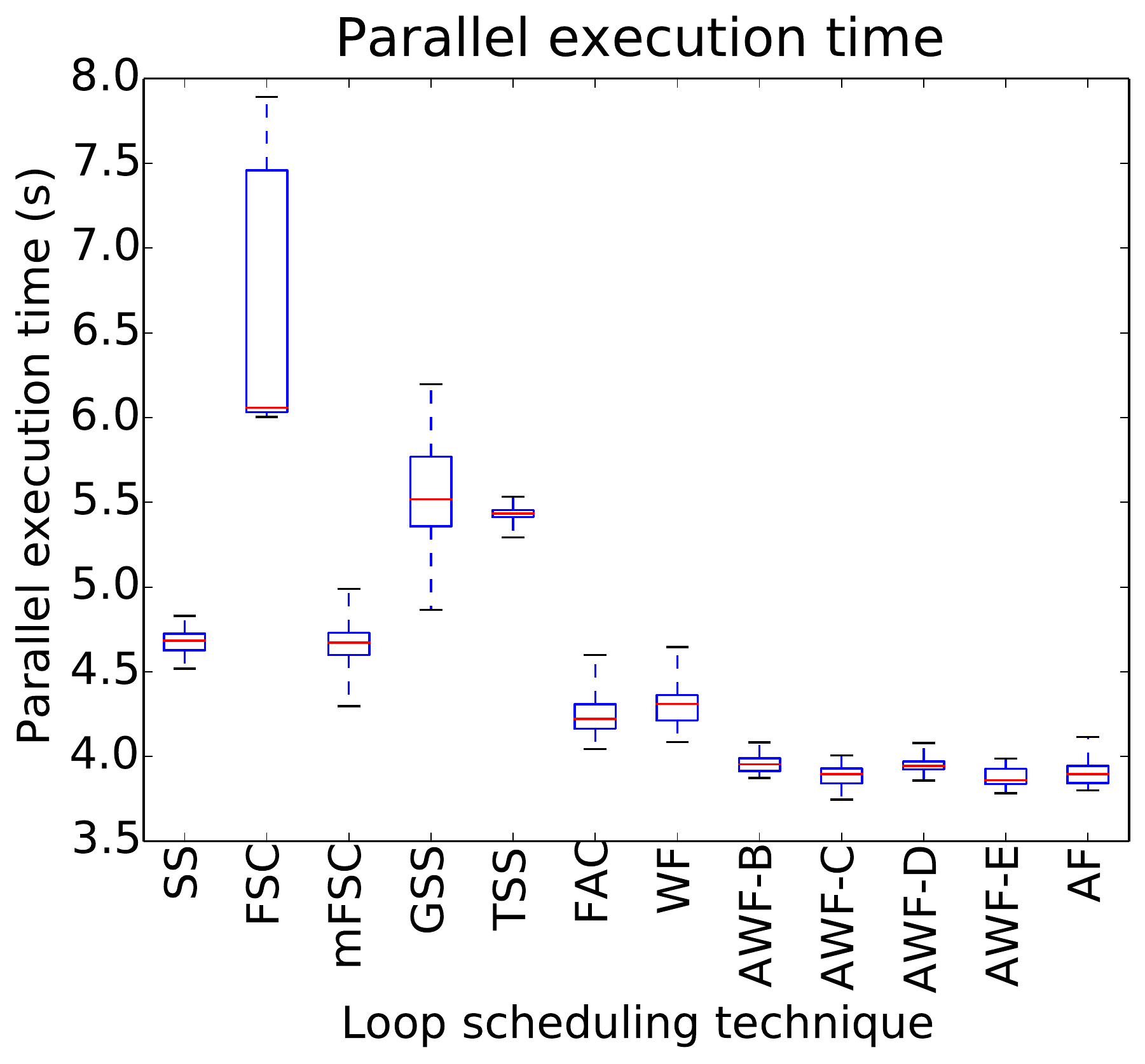}%
			\label{subfig:PSIA_PE_FT}%
		} \hspace{0cm} 
	\subfloat[PSIA - Network latency perturbations, robust]{%
		\includegraphics[clip, trim=0cm 0cm 0cm 0cm, scale=0.24]{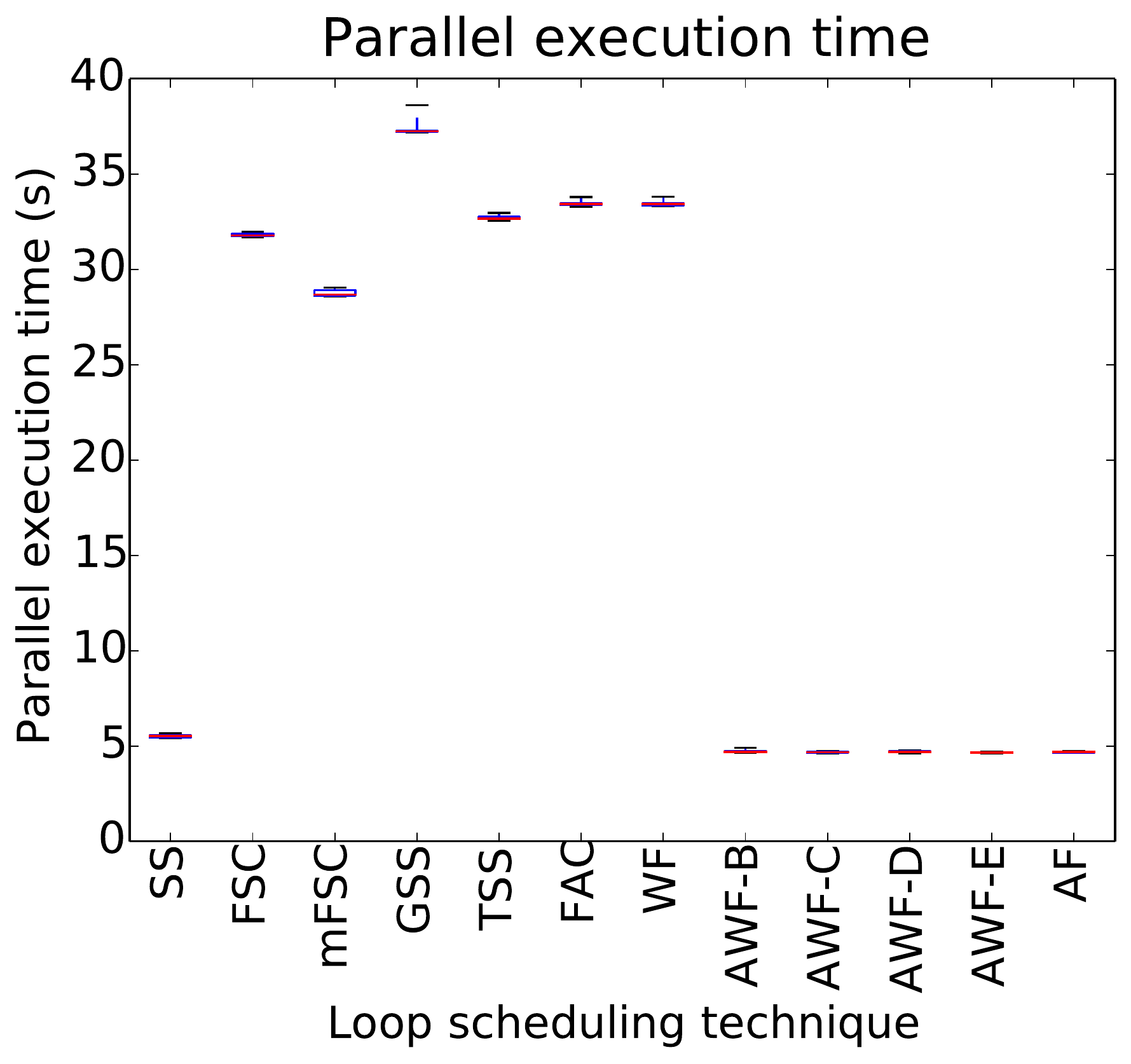}%
		\label{subfig:PSIA_lat_FT}%
	}  \hspace{0cm} 
\subfloat[PSIA - Combined PE and latency perturbations, robust]{%
	\includegraphics[clip, trim=0cm 0cm 0cm 0cm, scale=0.24]{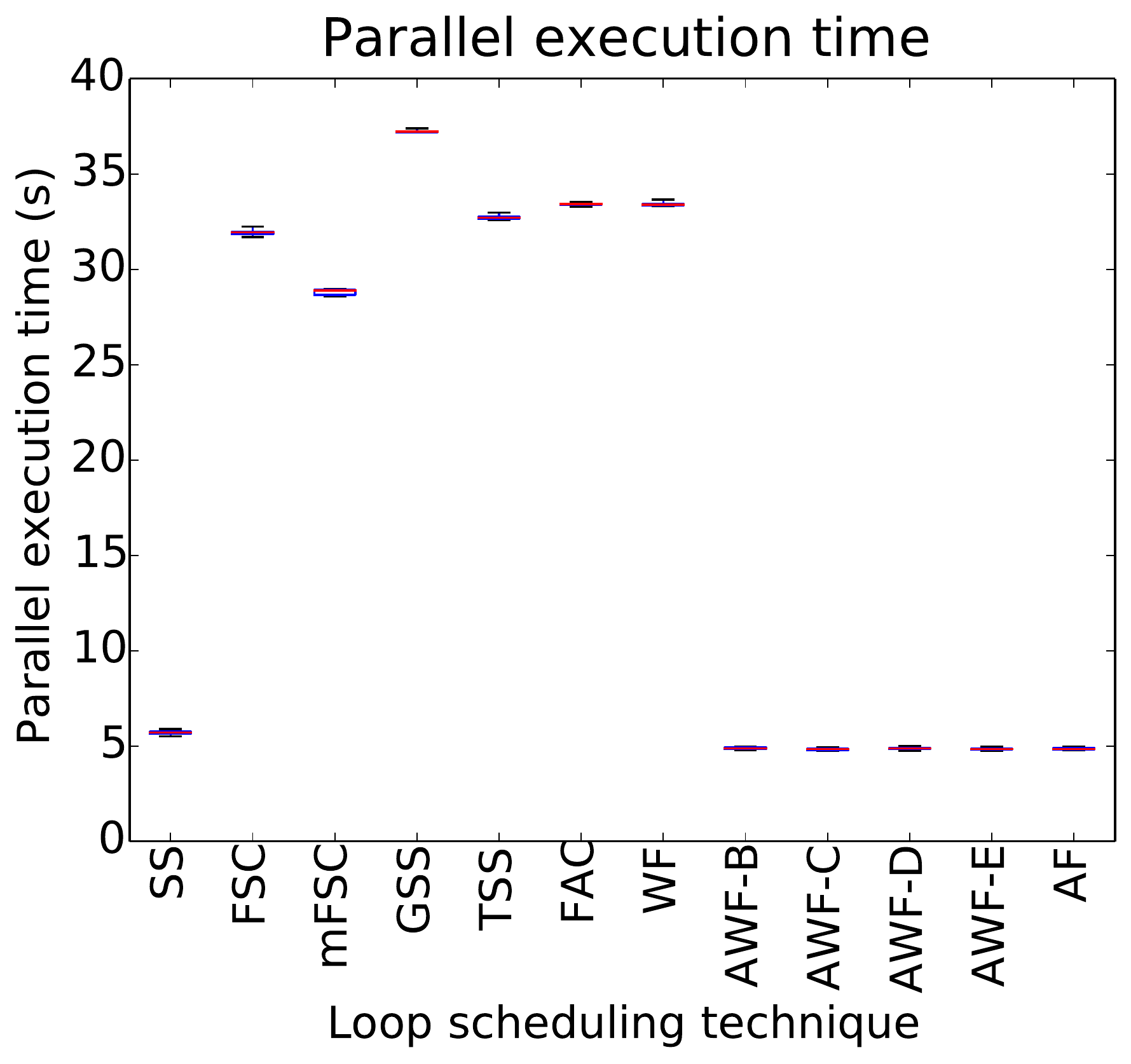}%
	\label{subfig:PSIA_pelat_FT}%
} \\		
		\caption{Performance of PSIA without and with robust \dlbTool{} under perturbations.}
		\label{fig:PSIA_perturbs}
	\end{minipage}
\end{figure*}

\begin{figure*}[htbp]
	\begin{minipage}{\textwidth}
		\centering
		\subfloat[Mandelbrot - PE perturbations, non robust]{%
			\includegraphics[clip, trim=0cm 0cm 0cm 0cm, scale=0.24]{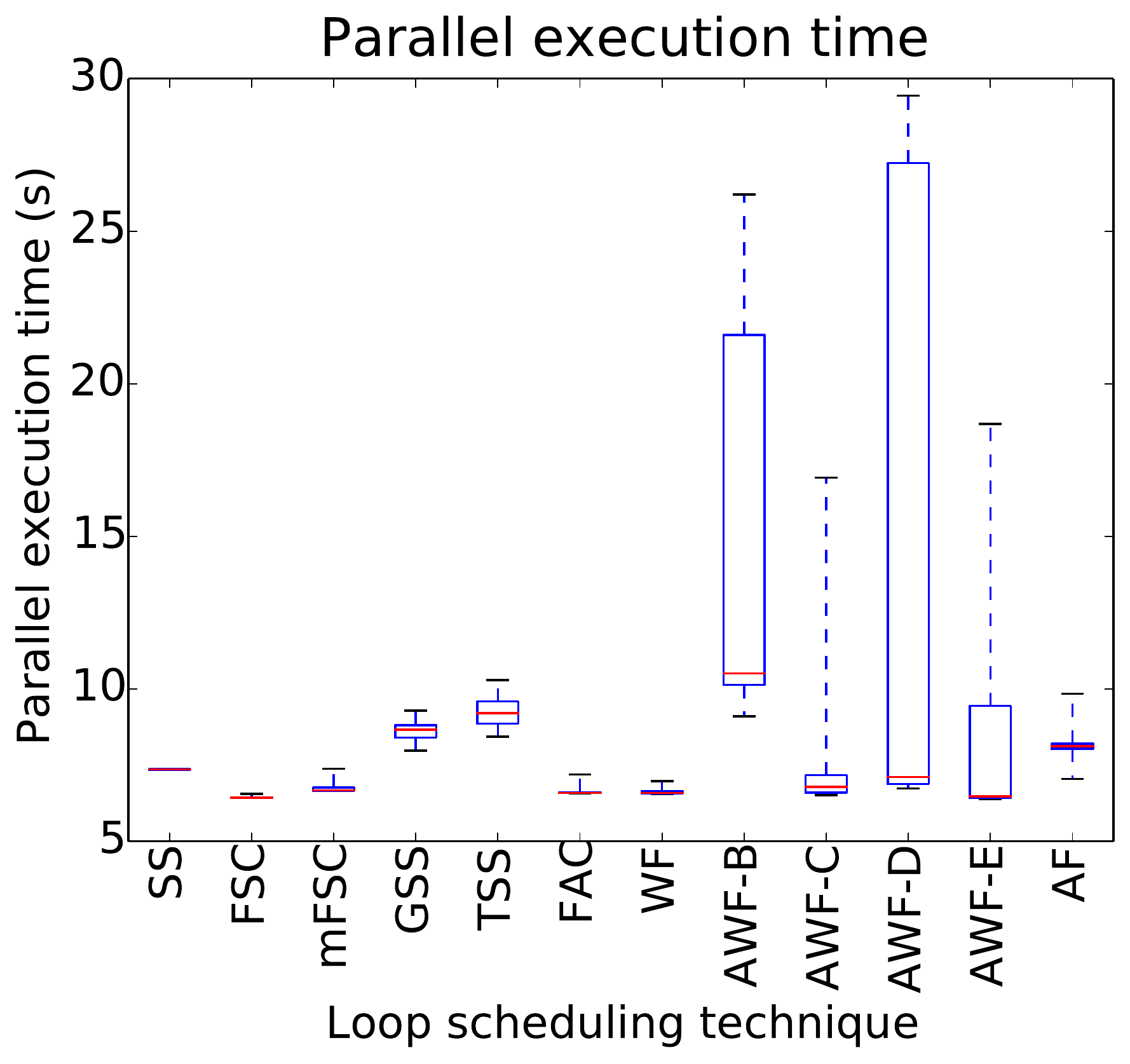}%
			\label{subfig:madel_PE_NOFT}%
		} \hspace{0cm} 
		\subfloat[Mandelbrot - Network latency perturbations, non robust]{%
			\includegraphics[clip, trim=0cm 0cm 0cm 0cm, scale=0.24]{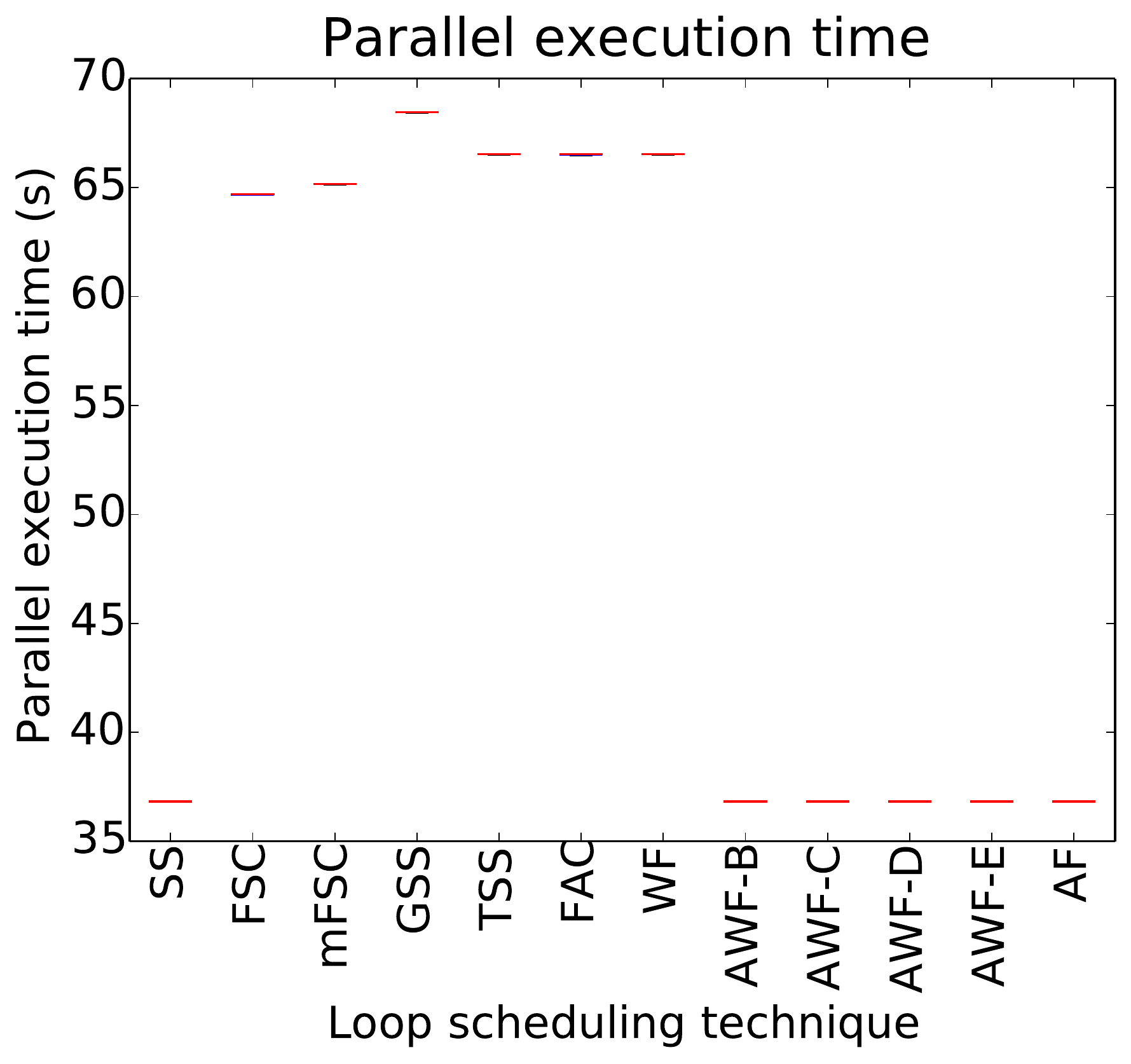}%
			\label{subfig:mandel_lat_NOFT}%
		} \hspace{0cm} 
    	\subfloat[Mandelbrot - Combined PE and latency perturbations, non robust]{%
		\includegraphics[clip, trim=0cm 0cm 0cm 0cm, scale=0.24]{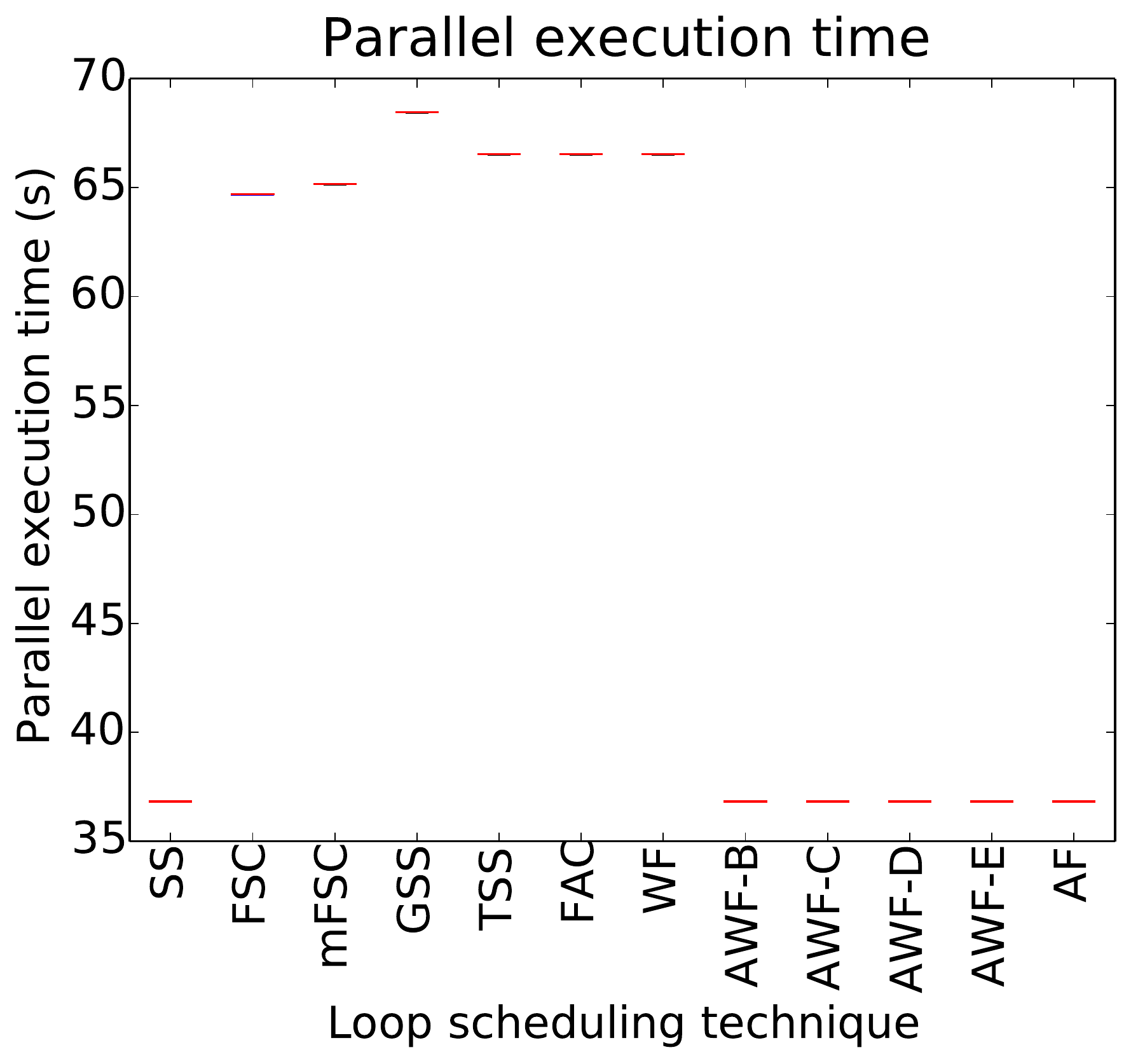}%
		\label{subfig:mandel_pelat_NOFT}%
	    }\\
	\subfloat[Mandelbrot - PE perturbations, robust]{%
		\includegraphics[clip, trim=0cm 0cm 0cm 0cm, scale=0.24]{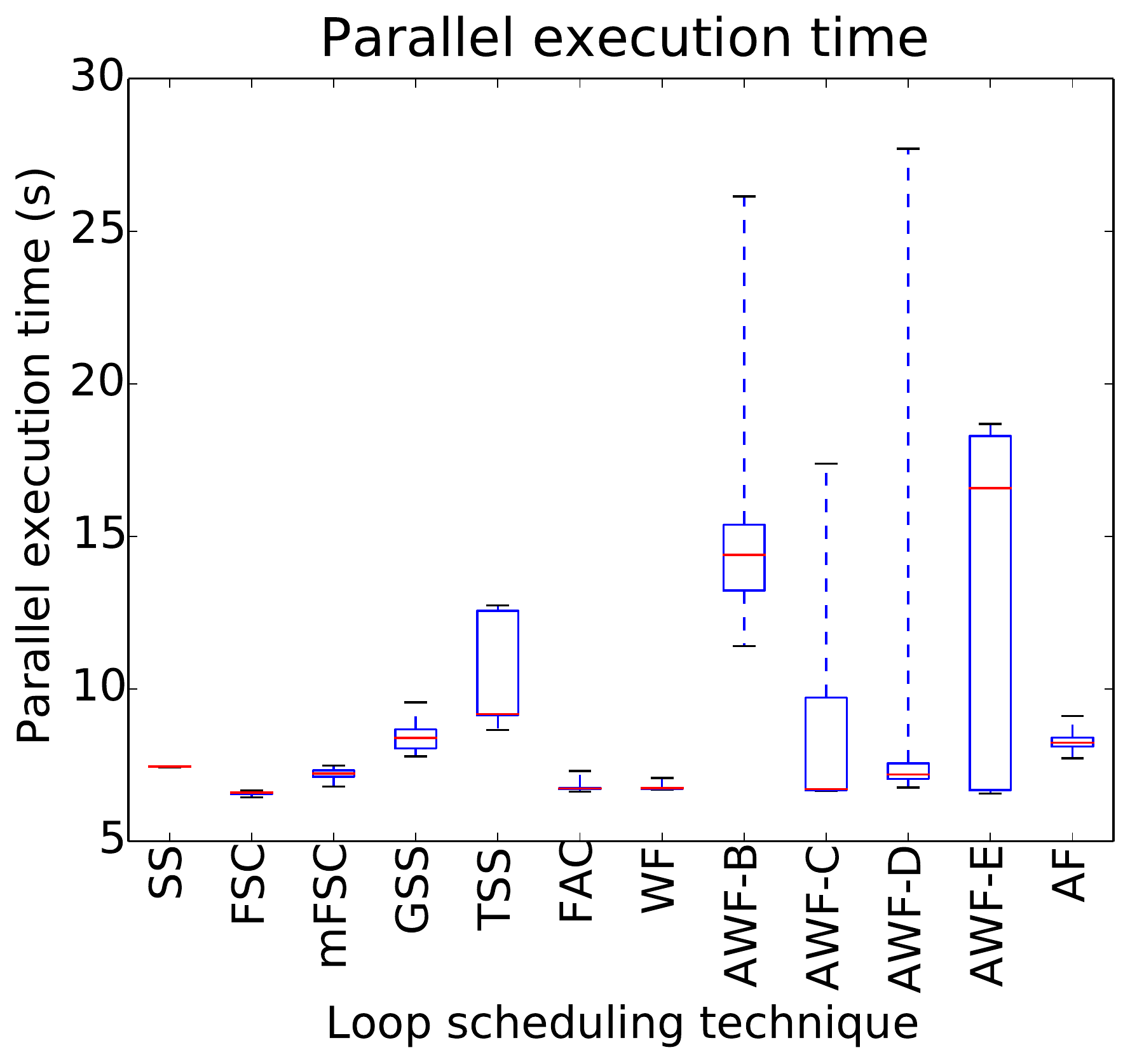}%
		\label{subfig:mandel_PE_FT}%
	}  \hspace{0cm}
	\subfloat[Mandelbrot - Network latency perturbations, robust]{%
	\includegraphics[clip, trim=0cm 0cm 0cm 0cm, scale=0.24]{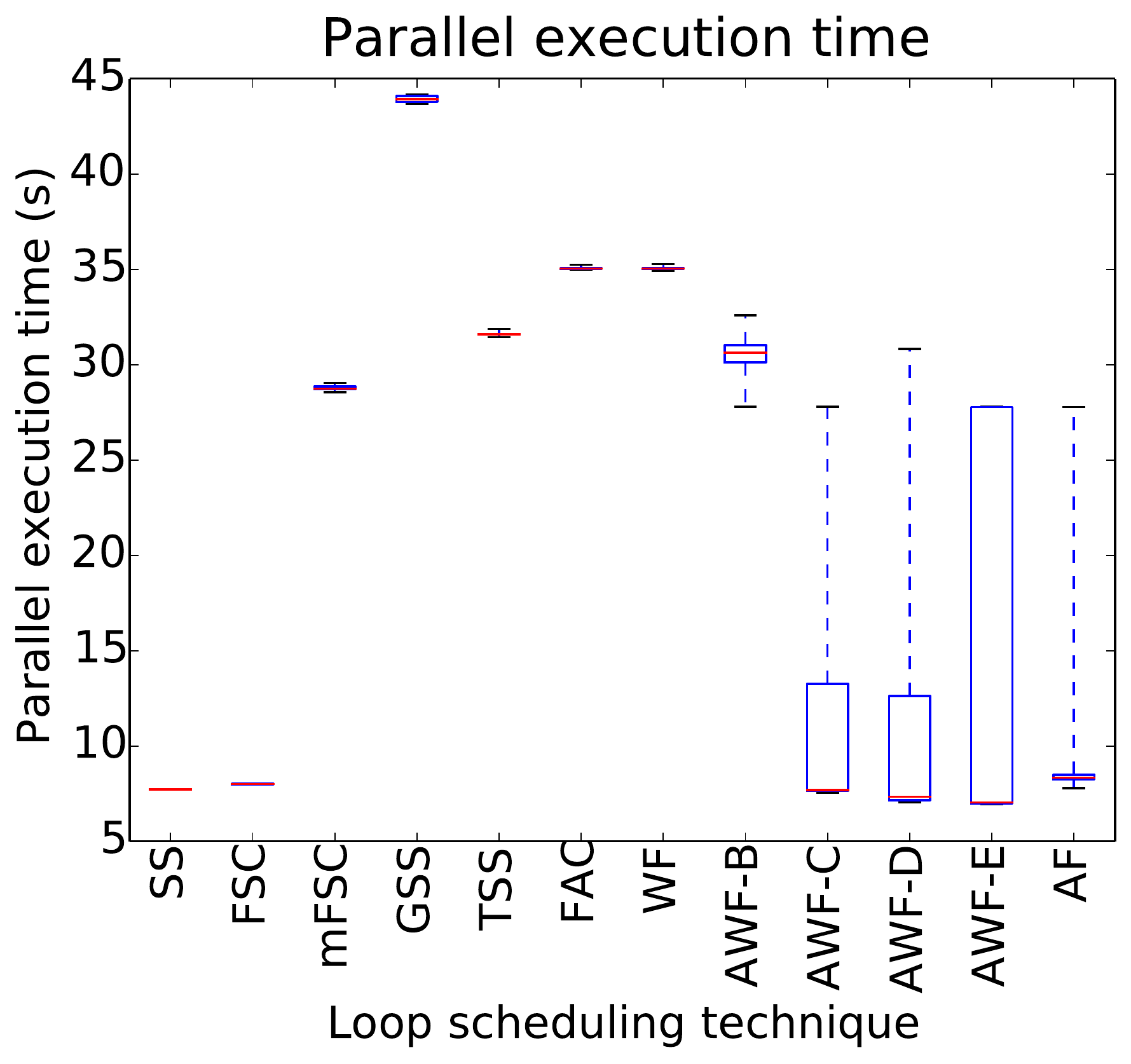}%
	\label{subfig:mandel_lat_FT}%
		}  
		\subfloat[Mandelbrot - Combined PE and latency perturbations, robust]{%
			\includegraphics[clip, trim=0cm 0cm 0cm 0cm, scale=0.24]{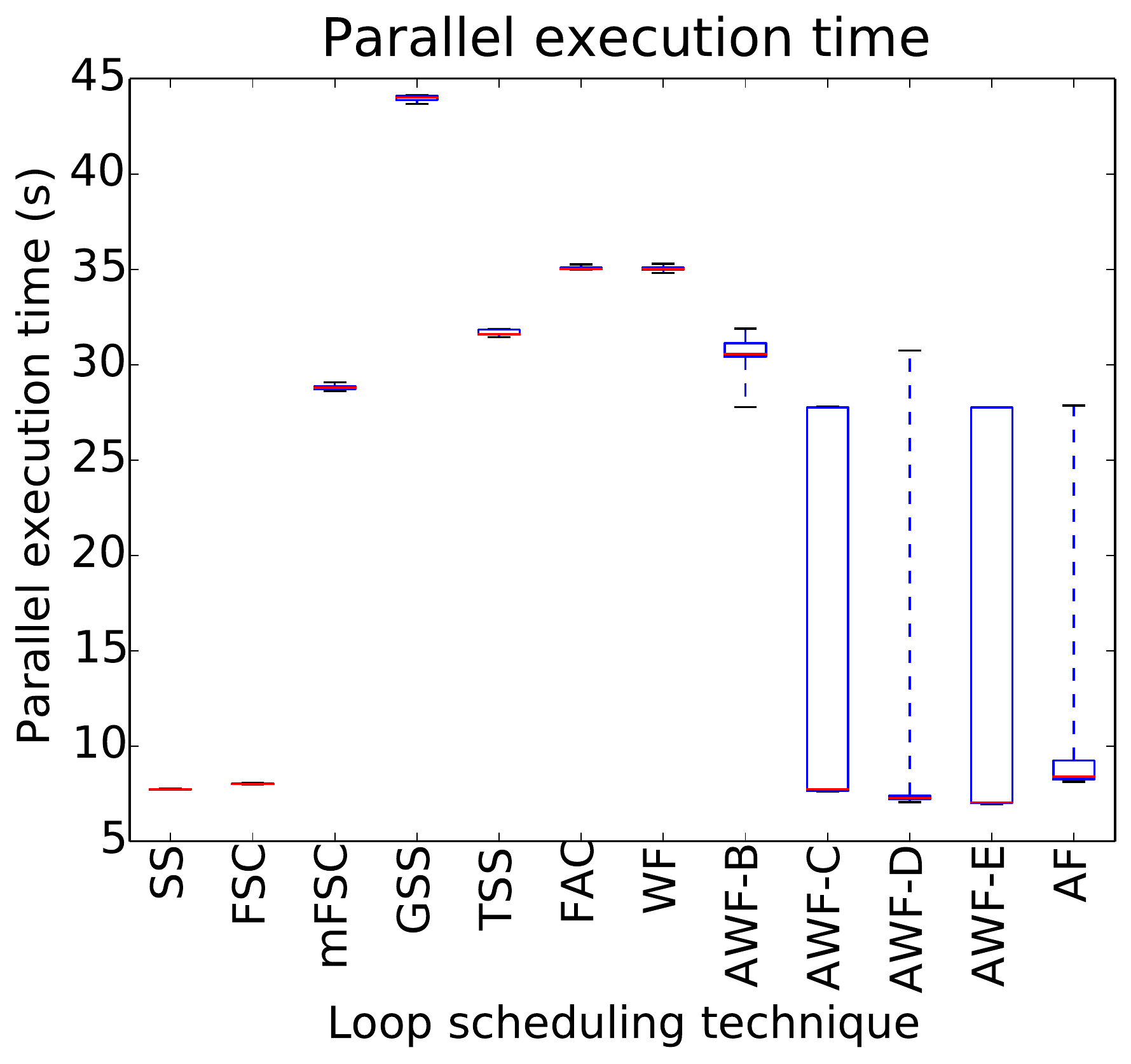}%
			\label{subfig:mandel_pelat_FT}%
		} \\		
		\caption{Performance of Mandelbrot without and with robust \dlbTool{} under perturbations.}
		\label{fig:Mandelbot_perturbs}
	\end{minipage}
\end{figure*}

%% file: 6.tex
\section{Conclusion and Future Work} 
\label{sec:conc}
A robust dynamic load balancing \rdlb{} is introduced in this work.
The theoretical analysis of the proposed approach shows its \discuss{linear scalability with the system size} and the quadratic decrease of its cost with increasing the number of PEs. 
The proposed \rdlb{} is integrated into the \dlbTool{} and used to schedule the loops of two \mbox{computationally-intensive} scientific applications under PE failures and perturbations.
Results show that \rdlb{}, applications tolerated up to ($P-1$ )PE failures with reasonable cost.
\rdlb{} boosted the robustness of DLS techniques in the case of execution under perturbations nearly $30$ times with adaptive DLS techniques.
DLS techniques with \rdlb{} achieved improved performance in terms of parallel execution time up to $7$ times faster compared to their counterparts without \rdlb{} in the presence of latency perturbations.

Extending \dlbTool{} to employ a decentralized control approach instead of a master-worker model and integrating \rdlb{} into the extended \dlbTool{} is planned in the future to avoid the master being a single point of failure.
The application of \rdlb{} to other independent task self-scheduling techniques is also envisioned in the future.
In addition, addressing silent data corruptions in large scale scientific applications is planned to ensure the integrity of applications results.
%

\section*{Acknowledgment}
This work has been partially supported by the Swiss Platform for Advanced Scientific Computing (PASC) project SPH-EXA: Optimizing Smooth Particle Hydrodynamics for Exascale Computing and by the Swiss National Science Foundation in the context of the ``Multi-level Scheduling in Large Scale High Performance Computers'' (MLS) grant, number 169123.